\documentclass[11pt]{article}

\usepackage{graphicx}
\usepackage{epsfig}
\usepackage{amsmath}
\usepackage{amsfonts}
\usepackage{cite}

\newcommand{\mys}[1]{\section{#1} \hspace{0.8cm}\setcounter{equation}{0}}
\renewcommand{\theequation}{\arabic{section}.\arabic{equation}}

\newcommand{\myappendix}{\appendix
   \renewcommand{\theequation}{\Alph{section}.\arabic{equation}}
   \vspace{30pt} \noindent {\Large \bf Appendices}}

\def\pslash{\hbox{/\kern-.5800em$p$}}

\def\gappeq{\mathrel{\rlap {\raise.5ex\hbox{$>$}}
{\lower.5ex\hbox{$\sim$}}}}
\def\lappeq{\mathrel{\rlap{\raise.5ex\hbox{$<$}}
{\lower.5ex\hbox{$\sim$}}}}
\newlength{\dummysp}
\newcommand{\Pf}{\mathop{{\hbox{Pf} \, }}\nolimits}

\newcommand{\beq}{\begin{eqnarray}}
\newcommand{\eeq}{\end{eqnarray}}

\newcommand{\e}{{\epsilon}}

\newcommand{\vev}[1]{{\langle #1 \rangle}}
\newcommand{\bigvev}[1]{{\left\langle #1 \right\rangle}}
\newcommand{\ord}[1]{{{\cal O}(#1)}}
\newcommand{\myref}[1]{(\ref{#1})}

\newcommand{\ben}{\begin{enumerate}}
\newcommand{\een}{\end{enumerate}}

\newcommand{\bit}{\begin{itemize}}
\newcommand{\eit}{\end{itemize}}

\newcommand{\Ocal}{{\cal O}}

\def\[{\left [}
\def\]{\right ]}
\def\({\left (}
\def\){\right )}

\begin{document}

\pagestyle{empty}

January 2007, revised July 2007
\hfill FTPI-MINN-07/01

\hfill UMN-TH-2534/07

\bigskip

\bigskip

\begin{center} 

{ \Large \bf  Chiral Lattice Gauge Theories and  The Strong 
  Coupling   Dynamics of a Yukawa-Higgs  Model  with Ginsparg-Wilson Fermions }

\bigskip

\bigskip

\bigskip

\bigskip

{\sc Joel Giedt}\footnote{Present address: Rensselaer Polytechnic Institute, Department of Physics, Applied Physics and Astronomy,
110 Eighth Street, Troy, NY 12180, USA}

\smallskip

{\it \small William I. Fine Theoretical Physics Institute\\
University of Minnesota\\
Minneapolis, MN 55455, USA}

\bigskip

and
\bigskip

 {\sc Erich Poppitz}

\smallskip

{\it \small Department of Physics\\ University of Toronto\\ Toronto, ON M5S~1A7, Canada}

\bigskip

{\tt giedtj@rpi.edu, poppitz@physics.utoronto.ca}

\vspace {1.5cm}

{\bf Abstract:} 
\end{center}
The Yukawa-Higgs/Ginsparg-Wilson-fermion construction of chiral
lattice gauge theories described in hep-lat/0605003 uses exact lattice
chirality to decouple the massless chiral fermions from a mirror sector, 
whose strong dynamics is conjectured to give cutoff-scale mass to the mirror
fermions without breaking the chiral gauge symmetry. In this paper, we study the mirror sector  dynamics of a  two-dimensional chiral gauge theory in the limit of strong Yukawa and vanishing gauge  couplings, in which case it  reduces to an $XY$ model coupled to Ginsparg-Wilson fermions.
For the mirror fermions to acquire cutoff-scale mass
it is believed to be important that the $XY$ model remain
in its ``high temperature'' phase, where there
is no algebraic ordering---a conjecture
supported by the results of our work.
We use analytic and Monte-Carlo methods with dynamical fermions to study the
scalar and fermion susceptibilities, and the mirror fermion
spectrum.  Our results provide convincing evidence that the strong dynamics
does not ``break" the chiral symmetry (more precisely, that the
mirror fermions do not induce algebraic
ordering in two-dimensions), and that 
the mirror fermions decouple from the infrared physics.

\vfill
\begin{flushleft}
\end{flushleft}
\eject
\pagestyle{empty}

\setcounter{page}{1}
\setcounter{footnote}{0}
\pagestyle{plain}

\section{Introduction and summary}

\subsection{Motivation}

The study of strong-coupling chiral gauge dynamics is an outstanding problem of great interest, both on its own and for its possible relevance to phenomenology.  Whereas the standard model of elementary particle physics is a weakly coupled chiral gauge theory, additional strong chiral gauge dynamics at (multi-) TeV scales may be responsible for breaking the electroweak symmetry and fermion mass generation.  Various tools are currently available for the study of the strong-coupling behavior of chiral gauge theories.
For instance, one has 't Hooft's anomaly matching and most attractive channel arguments.  These are
complemented by the ``power of holomorphy" in supersymmetric theories.  Scaling arguments and effective NJL-like models,
both using results from QCD as a stepping stone,
have also been employed extensively.
(Large-$N$ expansions, including the recently considered gravity
duals in the AdS/CFT (AdS/QCD) framework, do not usefully apply to
chiral gauge theories.)  None of these approaches
represents a ``first principles'' method, with an
accuracy that can be systematically improved.
Thus, the space-time lattice regularization
remains, to this day, the only way to advance
our limited knowledge of strong chiral gauge dynamics.

During the past two decades, since the work \cite{Ginsparg:1981bj} 
of Ginsparg and Wilson (GW), there has been significant progress 
in understanding chiral symmetries on the 
lattice \cite{Kaplan:1992bt, Narayanan:1993ss, Neuberger:1997fp,
Hasenfratz:1998ri, Luscher:1998pq} (further references 
are given in the reviews \cite{Golterman:2000hr, Luscher:2000hn}, 
while  \cite{Golterman:2004qv, Bhattacharya:2005xa} contain more 
recent work).
However, the lattice formulation of chiral gauge theories is still a
largely unsolved problem, although there
were some advances and promising two-dimensional
numerical results using the original
overlap Weyl determinant \cite{Narayanan:1996kz}.
The fact that a general formulation is lacking, 
along with the natural expectation that the lattice definition 
of a chiral gauge theory is not unique, indicates that the 
exploration of the various existing formulations, of their 
interconnections, as well as  of novel constructions 
of chiral lattice gauge theories are both warranted and worthwhile.

This paper is devoted to a study of a recently proposed
construction of lattice chiral gauge theories
\cite{Bhattacharya:2006dc}. It combines the GW exact lattice
chiral symmetry with earlier ideas of strong
coupling Higgs-Yukawa dynamics on the lattice, 
see \cite{Eichten:1985ft, Hasenfratz:1988vc,Stephanov:1990pc,
Golterman:1990zu,Golterman:1991re} 
and references therein. The essence of the proposal 
is that  the  exact lattice  chiral symmetry may
allow an old dream to  be realized: that the mirror
partners of the chiral massless fermions can be 
decoupled, without breaking the gauge symmetry.
The decoupling of the mirrors  is  possible in the strong-Yukawa symmetric phases of 
lattice fermion-Higgs models, where the fermions have
cutoff-scale mass without breaking the chiral symmetry.

The mechanism of mass generation without chiral symmetry breaking operative in strongly-coupled lattice theories has been known for some time:\footnote{In the two-dimensional  
context this mechanism was elucidated by Witten \cite{Witten:1978qu} via bosonization. In the case of four-dimensional strong four-Fermi interactions on the lattice, it  was  part of  the proposal of Eichten and Preskill  \cite{Eichten:1985ft} to formulate chiral gauge theories on the lattice. In the   context of strong-Yukawa lattice models it was discussed in \cite{Hasenfratz:1988vc,Stephanov:1990pc, Golterman:1990zu,Golterman:1991re}.} at strong Yukawa coupling, the charged mirror fermions form neutral bound states with the charged scalars; these bound states can pair up with other neutral mirror fermions, composite or elementary, to obtain mass without breaking the chiral symmetry.

The construction of \cite{Bhattacharya:2006dc}  gives an explicit 
definition of a local lattice action and measure. The global symmetry
Ward identities,  exact or anomalous,  are precisely as  in the target continuum 
theory, which  sets the proposal apart from earlier constructions aiming 
to decouple the mirror fermions. We believe that these desirable features 
are alone sufficient to motivate further study of the proposal.

The lattice model of \cite{Bhattacharya:2006dc} will  give  
rise to an unbroken chiral gauge theory only if there exists 
a strong-Yukawa  phase with an unbroken gauge symmetry,  
massless chiral  charged fermion spectrum,  and massive mirror
fermions. Because the construction makes use of the somewhat 
complicated exact lattice chiral symmetry, implemented via the 
Neuberger-Dirac operator \cite{Neuberger:1997fp},  
a strictly analytic approach---e.g., a strong-coupling 
expansion as used in \cite{Eichten:1985ft,Hasenfratz:1988vc,
Stephanov:1990pc, Golterman:1990zu}---to 
establishing the existence of the desired strong-coupling phase appears 
out of reach.  The purpose of this paper is to provide evidence, using
a semi-analytic approach that
computes the leading terms of the strong-coupling Yukawa expansion 
by Monte Carlo estimation, for the existence of the strong-coupling 
symmetric phase with a massive mirror fermion spectrum.

\subsection{Organization and summary of main results}

In Section \ref{theproposal}, we begin with a  review of the proposal  and of the arguments leading us to expect that   an unbroken gauge theory with a chiral spectrum of charged fermions emerges in the infrared.\footnote{See also Section IV in 
\cite{Bhattacharya:2006dc}.} We present the construction on  the example 
of a two-dimensional chiral $U(1)$ theory, the well-known ``345" theory. 
While the proposed lattice action can also be written in four  
dimensions, the simplicity of the analytical analysis,
and especially, the relative ease of the numerical analysis,
restricts the present  study of the dynamics to the two-dimensional case.

Of course, due to the special properties of two-dimensional
theories, the exact $U(1)$ symmetry cannot
be spontaneously broken \cite{MWC}.  However,
the gauge boson would acquire a large mass (relative
to the gap induced by the Schwinger mechanism) in the 
quasi-ordered phase, which is
analogous to spontaneous symmetry breaking
in the four-dimensional case.
It is for this reason that here, and
in several places below, we shall, 
with some abuse of language, use the 
term ``broken"  phase to refer to the
low-temperature phase of the two-dimensional 
$XY$ model, instead of the more appropriate 
algebraically- or quasi-ordered phase. 
Similarly, we will sometimes refer to the high-temperature 
disordered phase as the ``symmetric" or ``unbroken" 
phase and will take the liberty to omit the quotation marks.

In Section \ref{thetoymodel}, we formulate  a ``toy" version of the model, 
simple enough to be subjected to extensive 
analytical and numerical tests. The main focus of this paper is the Higgs-Yukawa dynamics of this ``toy" theory at strong Yukawa coupling and vanishing gauge coupling---the relevant limit to approach the continuum theory. The importance of this analysis is that it addresses precisely the question: is there a symmetric phase with a massless chiral spectrum of charged fermions in this model? A positive answer to this question is crucial for the viability of the proposal; a negative one would imply that  the proposal should be abandoned.

We begin the analysis in Section \ref{splitofZ}, where   we  show that the partition function of our model factorizes into a product of  ``light" and ``mirror" partition functions. 
We  work out the  splitting of the fermion measure into a product of positive and negative GW-chirality measures, using Neuberger's Dirac operator, and the resulting  factorization of the partition function; we give the technical details pertinent to this analysis  in Appendix \ref{Measuresplit}.
The factorization of the partition function is a unique feature of the GW-fermion Higgs-Yukawa theory  that makes it distinct from other Higgs-Yukawa approaches, 
notably the ``waveguide" models \cite{Kaplan:1992sg,Golterman:1993th,Golterman:1994at}.

The exact light/mirror split of the partition function at $g=0$ allows us to further concentrate entirely  on  the study of the dynamics of the mirror fermion-Higgs  theory.  The $g=0$ mirror fermion-Higgs theory of the ``toy" model is equivalent to the two-dimensional $XY$ model 
(see, e.g., \cite{id}) in the high-temperature disordered phase, $\kappa < \kappa_c$, deformed by the addition of fermions strongly  coupled to the spins. 
 
We next ask, in Section \ref{strongsymmetric}, whether  a strong coupling symmetric phase of the 
mirror-fermion Higgs-Yukawa theory exists.  We begin by noting 
that at strong Yukawa coupling the mirror fermion determinant 
is positive for arbitrary Higgs field background, 
provided the ratio of the Majorana to Dirac Yukawa 
couplings $h>1$.  The positivity of the fermion determinant 
permits us to numerically study the effect of 
mirror fermion ``loops'' on the Higgs field dynamics in the symmetric phase 
(i.e., at small Higgs kinetic term $\kappa$).\footnote{Actually,
``loops'' is a bit of a misnomer since our large Yukawa expansion
is just the opposite of the usual loop expansion.  Our
results are in that sense nonperturbative.}

We demonstrate the existence of a symmetric phase at strong Yukawa coupling by studying the scalar susceptibility in Section \ref{scalarsusc},  the  Binder cumulant---a quantity \cite{Binder:1981sa} probing higher-order correlations---in Section \ref{bindersection}, the vortex density in Section \ref{vortexsection}, and  the fermion bilinear susceptibilities in Section \ref{fermionsusc}.  
We use numerical simulations to investigate the behavior of  the various susceptibilities for different values of $\kappa$ and $h$  on lattices of size $N^2 = 4^2$, $8^2$, $16^2$. The Monte Carlo techniques we use are described
in Appendix \ref{simdetails}.
The combined study of the scalar and fermion bilinear order parameters and their scaling with $N$ presents ample 
evidence that  the theory is in the symmetric phase at strong Yukawa coupling to the  mirror fermions, provided $h>1$. 

For $h\leq 1$, still at small $\kappa$, we find evidence for criticality as appropriate in the low-temperature phase of the $XY$-model, where the   susceptibility  increases with the volume of the system; concurrently we find that the mirror fermion spectrum contains massless modes and that the vortex density decreases rapidly  as $h$ approaches unity from above, as in the 
Berezinski-Kosterlitz-Thouless transition to the low-temperature phase.

In summary, our results of  Section \ref{strongsymmetric} show  
that the strong coupling symmetric  phase exists and  thus 
the $U(1)$ symmetry---the global part of the gauge group---is unbroken. 
This result also implies that no fine-tuning is required 
in order to keep the theory in the symmetric phase  
(after gauging---keeping the gauge boson perturbatively massless\footnote{Setting
aside the effects of the Schwinger mechanism.}) 
at strong Yukawa coupling.  This  is in contrast with a naive perturbative argument which would, at large Yukawa coupling, imply a large backreaction of the mirror fermions on the Higgs dynamics, thus requiring fine-tuning of potentially many terms to keep the model in the symmetric phase. The small backreaction of the fermions on the Higgs dynamics at large Yukawa  is, however, consistent with the qualitative strong-coupling  arguments of Section \ref{theproposal} and \cite{Bhattacharya:2006dc}.

The next question we address is the  mirror fermion spectrum in the symmetric phase.
In Section \ref{chiralmassless}, we study the spectrum of both charged and neutral mirror fermions and show that they are all massive in the strong Yukawa symmetric phase, for    $h$   greater than unity. 

We note that, to the best of our knowledge, 
this is the first strong-Yukawa/Higgs construction
within the framework of the ``waveguide-like" 
(or Wilson-Yukawa) approaches to lattice chiral
gauge theories that successfully yields a chiral massless
spectrum of fermions at $g=0$.  Previous Yukawa
approaches have failed to yield a chiral light
spectrum of charged fermions already at the quenched, 
$g=0$, level, see \cite{Golterman:1993th, Golterman:1994at}
and references therein.  (The only exception that we know of is the quenched
study \cite{Hernandez:1997dd} of a theory that is in some respects similar
to our toy model, in the framework of the ``two-cutoff"
construction of \cite{HS}.  The advantages of
the model that we study here is that, unlike \cite{Hernandez:1997dd},
it exactly preserves the symmetries of the target theory
and the mirror fermions exactly decouple.)
In our construction, the crucial ingredient that leads
to a chiral spectrum is the decoupling, due to the exact
lattice chirality,  of the light chiral fermions from the strong 
Higgs-Yukawa dynamics of the mirror sector. It is the main feature
that distinguishes our models from earlier ``Wilson-Yukawa" or
``waveguide"  constructions.

\subsection{Conclusions and outlook} 

{\flushleft Our}   main  results  are: 
\begin{enumerate}
\item We have shown that exact lattice chirality can be used to decouple
the massless chiral fermions from a mirror sector. The light-mirror split of the fermion partition function is given in Appendix \ref{Measuresplit}.
\item We have given numerical evidence  that the strong
Yukawa dynamics of the mirror sector gives cutoff-scale
mass to the mirror fermions, without breaking the gauge symmetry (i.e. without inducing algebraic ordering in two dimensions).
\item We have found that, at strong Yukawa coupling,  the main effect of the mirror fermions
on the $XY$ model to which they are coupled is to renormalize
the hopping parameter $\kappa$ to smaller values---deeper into
the ``high temperature" phase---provided $h>1$.  For instance, in the $XY$ model coupled to fermions, the 
vortex density is higher, and the susceptibility is
lower, than in corresponding pure $XY$ model.  This conclusion
is further supported by the step function centered on
$h=1$ that develops in the Binder cumulant, extrapolated to the $N \gg 1$
limit.
\end{enumerate}
Our results show  that our proposed formulation of chiral lattice gauge theories  satisfies a first check on its viability: the strong Yukawa dynamics produces heavy mirrors without breaking the gauge symmetry.

\smallskip

{\flushleft{C}}learly, the results of this paper encourage further inquiry:
\begin{itemize}
\item
The toy model studied here gives rise to a would-be anomalous massless fermion spectrum in a trivial ($U=1$) gauge background. 
The issue of gauge anomalies in models where non-gauge dynamics in the mirror sector is introduced to decouple the mirrors was recently studied in \cite{Poppitz:2007ab}.
As explained there, in the anomalous case  a  result for the mirror spectrum at $U=1$ can not be used to infer the spectrum at $U\ne1$, since for an anomalous fermion content the mirror fermion partition function is not a smooth function exactly at $U=1$ and the split of the partition function into ``light" and ``mirror" is not consistent in this case (this is how the non-gauge mirror sector dynamics introduced to decouple mirrors is required to obey the anomaly free condition).  Clearly, the next important step is to study the mirror dynamics in an anomaly-free model, like the 345 model. See also the discussion in the Addendum of ref.~\cite{Poppitz:2007ab}.
\item
In the anomaly free case, we have reason to believe that our results will hold for nonvanishing gauge coupling, since the mirror-fermion/$XY$-model physics is at the ultraviolet scale, where the gauge coupling is weak. As shown in \cite{Poppitz:2007ab}, the mirror partition function is a smooth function of the gauge background in the anomaly free case and a result showing that the mirror fermions decouple at $U=1$ should persist for small nontrivial gauge background, e.g. in perturbation theory with respect to the gauge coupling. 
\item
A study analogous to the one of this paper in the four-dimensional case is  both feasible and desirable. We expect that  many of the details will be different---this is clear already from the fact that in four dimensions the  need for neutral massless ``spectator"   fermions, dictated  by two-dimensional Lorentz invariance, does not arise. We note in this regard the study of 
\cite{Gerhold:2006rc}, showing, within an analytic strong-coupling expansion the existence of a strong-coupling symmetric phase in a four-dimensional Yukawa-Higgs model with GW fermions (this result is  backed-up by Monte Carlo simulations \cite{GJ2}).  
\end{itemize}

\section{The proposal}

In this section, we review the proposal of \cite{Bhattacharya:2006dc}. It has many desirable features:  
\begin{itemize}
\item It is a full lattice proposal for a local action and measure for a chiral gauge theory.
\item  The realization of both   anomalous and anomaly-free global symmetries is exactly as in the target continuum theory.
\item  
Arguments in support of the conjecture that at strong 
Yukawa coupling the theory is in a symmetric phase 
with a chiral massless spectrum were given in 
 ref.~\cite{Bhattacharya:2006dc}.
\end{itemize}
We believe that the three points above warrant the  further study of the proposal. In this paper, we investigate the properties of this ``Yukawa-Higgs-GW-fermion" lattice model in more detail. First, we repeat the formulation of the model \cite{Bhattacharya:2006dc}. Then, we subject  the third point above  to a  detailed analysis.

\label{theproposal}

\subsection{The continuum ``345" model and its symmetries}
We will present the proposal on the example of a two-dimensional $U(1)$ chiral gauge theory. The condition for a two-dimensional $U(1)$ gauge theory to be free of gauge anomalies is that the sum of the squares of the charges of the left and right handed modes must cancel $
\sum_i q^2_{i,left} - \sum_j q^2_{j,right} = 0.$
An example of such a theory is the ``345" theory where there are left-handed fermions of charge $3$ and $4$ and right-handed fermions of charge $5$, so the chiral fermion spectrum is simply denoted as
$3_-, \, 4_-, \, 5_+$. The model is asymptotically free and can be solved  via bosonization (its spectrum  was found in \cite{Halliday:1985tg}).

The ``345" $U(1)$  theory has two anomaly free global symmetries, the ``133" symmetry (in a notation where $3_-$ has charge 1, and  $4_-$ and $5_+$---charge 3), and the global part of the ``345" gauge symmetry. The   fermion number ``111" symmetry has an anomaly and the associated 't Hooft operator is, schematically, of the form $(3_-)^3\; \partial_+ (4_-)^4 \; (\bar{5}_+)^5$.

\subsection{GW kinetic terms and exact chiral symmetries}
To begin describing the proposal, we introduce three two-dimensional Dirac fermions, $\Psi_{3}$, $\Psi_4$, $\Psi_5$, charged under the $U(1)$ gauge group with charges $3,4,5$, respectively. There is also a neutral Dirac fermion,  $\Psi_0$. The fermion fields live on the sites, labeled by $\{x\}$, of a two dimensional lattice and their lattice action consists of kinetic terms:
\begin{eqnarray}
\label{kinetic}
S_{kin} =\sum_{q = 0,3,4,5} \sum_{x,y} \bar\Psi_{q \;x} D_{q \; x,y} \Psi_{q \; y}~,
\end{eqnarray}
where $D_q$ is the Neuberger operator for a fermion of charge $q$, obeying the 
GW relation:\footnote{For a review of the GW relation and exact chiral symmetry on the lattice, see, for example, \cite{Luscher:2000hn} and references therein.}
\begin{equation}
\label{GW}
\{D_q,  \gamma_5 \} =   D_q \gamma_5 D_q ~.
\end{equation}
Here $\gamma_5$ is the appropriate two-dimensional matrix and the lattice spacing  has been set, from now on,  to unity.
The     lattice action (\ref{kinetic}) has a large number of exact global symmetries: 
\begin{eqnarray}
\label{kineticsymmetries}
\prod_{q = 0,3,4,5}  U(1)_{q, -} \times U(1)_{q,+}~,
\end{eqnarray}
where $U(1)_{q,\pm}$ acts only on the Dirac fermion field of charge $q$:
\begin{eqnarray}
\label{kineticsymmetries2}
\Psi_q \rightarrow e^{i \alpha_{q, \pm} P_\pm} \Psi_q ~~,~~ 
\bar\Psi_q \rightarrow \bar\Psi_q e^ {-i\alpha_{q, \pm} \hat{P}_\mp}~,
\end{eqnarray}
where $P_\pm = (1 \pm \gamma_5)/2$ and $\hat{P}_\pm = (1 \pm \hat{\gamma}_5)/2$, with
 $\hat{\gamma}_5 \equiv (1 - D) \gamma_5$. That $\hat{\gamma}_5^2 = 1$ follows from the GW relation (\ref{GW}); note that  $\Psi_q$ and $\bar\Psi_q$ transform differently, which is perfectly natural in Euclidean space. The projector used for every $\Psi_q$ involves the appropriate Neuberger operator $D_q$.

That the symmetries in equation~(\ref{kineticsymmetries2}) are all exact
symmetries of the action (\ref{kinetic}) follows  from $\hat{P}_\mp D = D P_\pm$, yet another consequence of the GW relation (\ref{GW}). The measure of integration, however, is not invariant under any individual $U(1)_{q, +}$ or $U(1)_{q, -}$. Instead, under an infinitesimal   $U(1)_{q, \pm}$ transformation (\ref{kineticsymmetries2}) with parameter $\alpha_{q, \pm}$, the measure  changes as follows:
\begin{eqnarray}
\label{changeofmeasure} 
& &U(1)_{q, \pm}: \prod_{r = 0,3,4,5} d\bar\Psi_r d\Psi_r  \rightarrow   \\
 \prod\limits_{r = 0,3,4,5} & d\bar\Psi_r d\Psi_r  &\left[1 -i \alpha_{q, \pm} {\rm Tr} \left( P_\pm - \hat{P}_{\mp}   \right)\right]    = \prod\limits_{r = 0,3,4,5} d\bar\Psi_r d\Psi_r  \left[1 \pm i \alpha_{q, \mp} \frac{1}{2} {\rm Tr}  \; D_q  \gamma_5 \right]. \nonumber
\end{eqnarray}
Eqn.~(\ref{changeofmeasure}) implies that 
for vectorlike symmetries $U(1)_{qV}$ ($\alpha_{q,+} = \alpha_{q, -}$),  there is no Jacobian and thus they are true symmetries of the theory. On the 
other hand \cite{Hasenfratz:1998ri,Fujikawa:1998if}, since $- \frac{1}{2} {\rm Tr}   D_q  \gamma_5  = n_+^0 - n_-^0$, the difference between the number of  left- and right-handed zero modes of $D_q$, the continuum violation of charge for anomalous symmetries is reproduced  by the nonzero Jacobian.

\subsection{Reduction of the global chiral symmetries by GW-Yukawa couplings}

To construct our candidate ``345" chiral lattice theory, we introduce a unitary higgs field, $\phi_x$,  living on the lattice sites.   We will  use $\phi_x$ to write all possible Dirac and Majorana Yukawa  terms that violate all symmetries (\ref{kineticsymmetries}) of the kinetic term  (\ref{kinetic}) except:
\begin{equation}
\label{masssymmetries}
U(1)_{3,-}\times U(1)_{4,-} \times U(1)_{5,+} \times U(1)_{0, +}~.
\end{equation}
 To this end, we relate the Dirac fields $\Psi_q$ to their  chiral components: $\Psi_{q, \pm} \equiv P_{\pm} \Psi_q$, $\bar\Psi_{q,\pm} \equiv \bar\Psi_{q} \hat{P}_{\mp}$; note that the definition of the   $\bar\Psi_\pm$ chiral  modes is now both momentum and gauge-background dependent. 
We then write down the most general Dirac and Majorana couplings---expected to give mass  of order the inverse cutoff---to the fields:
\begin{eqnarray}
\label{xandy}
X_+ &=& (\Psi_{3,+}^T \: \bar\Psi_{3,+} \: \Psi_{4,+}^T \: \bar\Psi_{4,+}) \nonumber \\
Y_- &=& \left(\begin{array}{c} \Psi_{5,-} \cr \bar\Psi_{5,-}^T \cr \Psi_{0,-}\cr \bar\Psi_{0,-}^T\end{array}\right)~,
\end{eqnarray}
where $T$ denotes transposition (we treat unbarred Dirac spinors as columns and barred ones as rows) of the form:\footnote{The presence of 
neutral mirrors is necessary to preserve two dimensional Lorentz invariance.}
\begin{eqnarray}
\label{DMmass}
S_{mass} = y \sum_{x} X_{+ \;x} \; M(\phi_x,\phi^*_x) \; Y_{- \; x}~.
\end{eqnarray}
Both Dirac and Majorana masses are  to be included for the mirror fields: if Majorana masses are omitted, there will be extra unbroken chiral symmetries and unlifted zero modes in an instanton background, resulting in failure to reproduce  the 't Hooft vertex. Therefore, Majorana masses are needed in order to introduce a violation of fermion number into the lattice, as  has already been noted in \cite{Golterman:2002ns}.

Instead of writing explicitly the entire matrix $M(\phi_x, \phi_x^*)$, which depends on powers of $\phi$ and $\phi^*$ determined by gauge invariance, as indicated in (\ref{DMmass}), we give an example of  a Dirac mass term: $\bar\Psi_{0,-} (\phi^{*})^3 \Psi_{3,+}  + \bar\Psi_{3,+} \phi^3\Psi_{0,-} $,  and of  a Majorana mass of the form: $\bar\Psi_{5,-} \gamma_2 \phi^8 (\bar\Psi_{3,+})^T - \Psi_{3,+}^T  \gamma_2 (\phi^{*})^8 \Psi_{5,-}$. Here $\gamma_2$ is the hermitean   gamma matrix that appears when Majorana masses are written using Dirac spinors, while $\phi$ is the unitary Higgs field. Thus, the explicit form of $M(\phi_x, \phi_x^*)$ in (\ref{DMmass}) contains appropriate powers of $\phi$ and $\gamma_2$-insertions.
The general mass matrix (\ref{DMmass}) violates all $U(1)$ 
symmetries except the desired $U(1)_{3,-}\times U(1)_{4,-} \times U(1)_{5,+} \times U(1)_{0, +}$ 
symmetry  from(\ref{masssymmetries}).

We continue by noting that not all symmetries (\ref{masssymmetries}) of  the lattice action (\ref{DMmass}), (\ref{kinetic}) are respected by the lattice path integral measure.  The measure is only invariant under  three symmetries: the $U(1)_{345}$  and the $U(1)_{133}$ chiral symmetries---linear combinations of  $U(1)_{3,-} \times U(1)_{4,-} \times U(1)_{5,+} $ with coefficients $345$ and $133$, respectively---and the $U(1)_{0,+}$, which acts only on the $n_+ \equiv P_+ \Psi_{0}$ neutral field, whose dynamics is expected to decouple from the physics of the charged sector.
The third linear combination of the first three $U(1)$'s in equation~(\ref{masssymmetries})---the fermion number symmetry of the light charged fields, which can be taken to be the ``$111$'' symmetry---has an anomaly exactly reproduced by the Jacobian, eqn.~(\ref{changeofmeasure}), of the corresponding transformation  of the measure; see  \cite{Hasenfratz:1998ri}, 
\cite{Fujikawa:1998if}, and references in \cite{Luscher:2000hn}. Thus, the lattice theory obeys exact Ward identities, including the anomalous ones. For example, using (\ref{changeofmeasure}) one finds  that  the 111 transform of an operator $\cal{O}$ obeys the exact lattice Ward identity: 
\begin{eqnarray}
\label{WI}
\langle \delta_{\alpha_{111}}  {\cal{O}} \rangle =
i\; \frac{\alpha}{2} \; \langle  {\cal{O}} \; {\rm Tr}\left[ \gamma_5 (D_3 + D_4 - D_5)  \right]  \rangle~.
\end{eqnarray}
The continuum limit expansion  Tr$\gamma_5 D_q \sim  \int d^2 x \; \epsilon^{\mu \nu} F_{\mu \nu}$  \cite{Fujikawa:1998if} implies that the anomalous Ward identity (\ref{WI}) has 
the expected continuum limit.

At the end of this section, it is worth noting that this proposal carries some
of the flavor\footnote{We thank David B. Kaplan for pointing this out to us. We also note that a proposal to decouple the mirrors by combining the (approximate) lattice chirality of domain wall fermions with the Eichten-Preskill ideas  was made earlier in \cite{Creutz:1996xc}.}  of an earlier construction of Eichten and Preskill
\cite{Eichten:1985ft}, attempting to use strong four-Fermi interactions to decouple mirrors and doublers (it is clear that integrating out our  short-ranged $\phi_x$ will produce strong multi-fermion interactions of the mirrors). 
Their proposal is  known not to give rise to a chiral gauge
theory (see  \cite{Golterman:1992yh}, where the similarity with 
Yukawa models was also used). In our case, the modified lattice
chiral symmetry that leads to exact decoupling of the chirality 
components only allows us to make use of the Yukawa analogue of the strong four-Fermi coupling symmetric phase (see the Appendix of \cite{Eichten:1985ft})---a phase with unbroken gauge symmetry, where all fermions that participate in the strong interactions are massive. 

\subsection{Action, partition function,  and dynamics}

To ensure that the dynamics of our lattice model reproduces that of the desired unbroken chiral gauge theory, 
we need to demonstrate the existence of a strong-Yukawa-coupling symmetric phase with chiral spectrum of massless fermions (recall again the strong coupling 
analysis of \cite{Golterman:1994at} which showed that in the waveguide model the spectrum in this phase was vectorlike). 
 Remarkably, as we will find below,  to leading order in the strong Yukawa coupling expansion and small gauge coupling---precisely the regime where the waveguide idea broke down---there appear no new massless modes and the spectrum of the unbroken gauge theory is  chiral. 
 
 The total action of the lattice model is, finally: 
\begin{eqnarray}
\label{totalL}
S  =  S_{Wilson} + S_{kin} + S_{mass} + S_\kappa~, 
\end{eqnarray}
$S_{kin}$ is defined in (\ref{kinetic}), $S_{mass}$---in (\ref{DMmass}), $S_{Wilson}$ is the usual plaquette action for the link variables $U_{x, x+ \hat\mu}$,  appropriately modified to restrict the gauge field path integral to  admissible gauge field backgrounds, see \cite{Luscher:2000hn}, and 
 $S_\kappa$ is the action for the charge-1 unitary Higgs field:
\begin{eqnarray}
\label{Skappa}
S_\kappa =  \frac{\kappa}{2}\; \sum_{x} \sum\limits_{\hat{\mu}} \left[ 2 - \left(\; \phi_x^* \; U_{x, x+ \hat\mu} \; \phi_{x+\hat\mu} + {\rm h.c.}\; \right) \right]~.
\end{eqnarray}
The dynamical issue that needs to be addressed is the existence of an ``unbroken''
phase where $\phi$ is disordered (analogous to $\vev{\phi}=0$,
versus $\vev{\phi} \not= 0$, in four dimensions), such that the gauge boson is massless. 

In the case without fermions, it is well known  \cite{Fradkin:1978dv}  that theories with unitary Higgs fields (contrary to ``everyday" continuum intuition) exhibit a symmetric phase, for small enough $\kappa$.
The essential idea\footnote{Sometimes called the ``FNN mechanism" \cite{Forster:1980dg}.}   is that for small $\kappa$  large fluctuations of the unitary Higgs field---or, in the equivalent unitary gauge, the pure-gauge fluctuations of the gauge field $U$---are not suppressed by the action (\ref{Skappa}) and hence their   correlation length  is of order the lattice spacing. Thus, integrating out the rapidly  fluctuating Higgs fields results in   renormalization of the gauge coupling plus a tower of higher-dimensional gauge invariant local operators which are irrelevant for the long-distance physics of the gauge theory.  This is most easily seen upon  integrating over the rapid fluctuations of $\phi_x$, or equivalently, the pure-gauge part of $U$, by explicitly performing the strong-coupling (small $\kappa$) expansion. The  leading  correction is a small, ${\cal{O}}(\kappa^4)$, shift  to the inverse gauge coupling constant, $g^{-2} \rightarrow g^{-2} + \kappa^4$ (see section II.C.(a) of \cite{Fradkin:1978dv}), while the tower of   higher-dimensional gauge-field dependent operators is suppressed by increasing powers of the small correlation length of $\phi_x$.

Since the gauge theory (\ref{totalL}) is asymptotically free, to study the continuum limit it is sufficient to begin at leading order in the $g \rightarrow 0$ expansion;  here we will confine our analysis to this limit. The limit  freezes the gauge degrees of freedom to $U=1$; see also discussion in Section \ref{splitofZ}. The resulting theory is a unitary Higgs-Yukawa model, equivalent to the $XY$ model coupled to fermions, whose phase structure can be studied in various limits.  We are interested in the symmetric phase of the lattice $XY$ model and will  study $\kappa < \kappa_c$  while also taking   $y \rightarrow \infty$ (cf. eqn.~(\ref{DMmass})).

In a trivial gauge background, the lattice partition function of the model (\ref{totalL}) factorizes 
into a product
$Z = Z_{light} \times Z_{mirror}$, with:
\begin{eqnarray}
\label{Zmirror}
Z_{light} &=& \int \prod_{x} d \Psi_{3,-} d \Psi_{4,-} d \Psi_{5,+}d \Psi_{0,+}\;  e^{- S_{kin}[\Psi^{light}]} ~, \\
 Z_{mirror}&=& \int \prod_{x} d \Psi_{3,+} d \Psi_{4,+} d \Psi_{5,-}d \Psi_{0,-} d \phi \;     e^{- S_{kin}^{mirror}[\Psi^{mirror}] - S_\kappa[\phi] - S_{mass}[\Psi^{mirror}, \phi] } ~, \nonumber
\end{eqnarray}
where we used the decomposition of the Dirac fermion measure into left- and right-chirality measures, $d \Psi= d \Psi_+ d \Psi_-$, see following Sections, and, 
for conciseness,    omitted the conjugate fields in the measure. We also denoted collectively by $\Psi_{light}$ the fields $\Psi_{3,-}, \Psi_{4,-}, \Psi_{5,+}$, $\Psi_{0,+}$, and by $\Psi_{mirror}$ the heavy charged mirrors $\Psi_{3,+}, \Psi_{4,+}, \Psi_{5,-}$, and the neutral $\Psi_{0,-}$. The ``mass" term is given by equation~(\ref{DMmass}) and  the kinetic term for $\phi$ by~(\ref{Skappa}). 

The most important point is the splitting of the kinetic terms (\ref{kinetic}) into  light and mirror modes in (\ref{Zmirror}). This  follows from the identity, which also holds in an  arbitrary gauge background:
\begin{eqnarray}
\label{kinetic2}
\bar\Psi_{q}  D_q  \Psi_q = \bar\Psi_{q, +} D_q \Psi_{q, +} + \bar\Psi_{q,-} D_q \Psi_{q,-}~,
\end{eqnarray}
where the   cross-terms vanish due to the GW relation (\ref{GW}). Thus the mirror and light partition functions factorize at $g=0$.    We stress that the GW relation was crucial in order for (\ref{kinetic2}) to hold; there is no other way to achieve (\ref{kinetic2}), and hence the factorization (\ref{Zmirror}), on the lattice, for $D_q$ free of doublers.  

Finally, let us study $Z_{mirror}$ and its effect on the light modes, in the $y \rightarrow \infty$ and $\kappa < \kappa_c$ limit. Of particular concern is the possible appearance of extra massless states and the associated vanishing of the mirror determinant. 
To this end, we redefine the mirror fermion fields in (\ref{Zmirror}), $\Psi_{3,+}, \Psi_{4,+}, \Psi_{5,-}$, and the singlet $\Psi_{0,-}$, by $1/\sqrt{y}$. This multiplies their kinetic terms by $1/y$. Thus, as $y \rightarrow \infty$, the mirror fields kinetic terms vanish, and the mirror  fermion action consists solely of a Yukawa term given by (\ref{DMmass}) with $y = 1$:
\begin{eqnarray}
\label{DMmirror}
S_{mass} =  \sum_{x} X_{+ \;x} \; M(\phi_x,\phi^*_x) \; Y_{- \; x}~.
\end{eqnarray}
We can now perform the  integral over the mirror fermions in $Z_{mirror}$. 
The notation in the above equation might lead one to believe that the mirror fermion path integral is a product of strictly local factors:
$$
\prod\limits_{x} {\rm det} M(\phi_x,\phi^*_x)~,  $$ 
where $M(\phi_x,\phi^*_x)$ is the $8 \times 8$ matrix from (\ref{DMmass},\ref{DMmirror}), which is gauge covariant and depends only on the values of $\phi_x$ at $x$. Since $\phi_x$ is unitary, this would imply  that the gauge invariant ${\rm det} M(\phi_x,\phi^*_x)$ is {\it a.)} $\phi$-independent and {\it b.)} nonzero.
Hence, we would have been led to believe that the dynamics governing the  fluctuations of the pure-gauge degrees of freedom is unaffected by  the mirror fermions, to leading order in $1/y$. Moreover, this argument would also imply that there is no fine-tuning, at large $y$, required in order to keep the $XY$ model in its  high-temperature phase. Finally, a constant  determinant, as would be obtained at large-$y$ from the above argument, 
indicates that there are no massless fermion states, as a massless fermion state is expected to lead to a  zero determinant.
 
The true story, however, is more complicated than the discussion of the previous paragraph. This is due to the fact that the various
$\bar\Psi_{  \pm}$  chiral components which enter  $X_{+ \; x}, Y_{- \; x}$ (\ref{xandy}) and $S_{mass}$ are somewhat smeared due to the nonlocality of the chiral projectors that  define the chiral components for the barred fields. However, the extent of the nonlocality of  the $\bar\Psi_\pm$-component is small, governed by the range of nonlocality of  Neuberger's operator, which is of order of the lattice spacing with an exponential tail, as the analysis of   \cite{Hernandez:1998et, Neuberger:1999pz} shows.  Hence, one expects that  the qualitative arguments of the previous paragraph still hold, together with the conclusion that the mirror fermion fluctuations do not significantly affect the pure gauge fluctuations and their determinant is nonzero. Section \ref{strongsymmetric} is devoted to verifying this conjecture.

\section{The simpler ``toy" model}
\label{thetoymodel}

\subsection{Definition of the toy model: action and symmetries}
\label{deftoy}

In our analytic and numerical study, we will use a  simpler model that captures the main features of the mirror sector dynamics at $g=0$. The model has a minimal field content, allowing an exhaustive
study of the phase diagram using numerical methods with the  computer resources available to us. 
 
Our toy model is a $U(1)$ lattice gauge theory with one charged Dirac fermion, $\psi$, of charge 1, and a neutral spectator, $\chi$. 
 The desired spectrum of light fields in the target theory is the charged $\psi_+$ and the neutral $\chi_-$. 
 The chirality components for the charged and neutral fermions are defined, as in the previous section, by the projectors which include  the appropriate Neuberger operators $D_1$ or $D_0$ for the barred components.
  The fermion part of the action of our toy model is:
\begin{eqnarray}
\label{toymodel}
S&=& S_{light} + S_{mirror} \\
S_{light} &=&
 \left( \bar\psi_+, D_1 \psi_+\right) + \left( \bar\chi_-, D_0 \chi_-\right) \nonumber  \\
 S_{mirror} &=& \left( \bar\psi_-, D_1 \psi_-\right) + \left( \bar\chi_+, D_0 \chi_+\right)  \nonumber \\
&+& y \left\{ \left( \bar\psi_- , \phi^* \chi_+ \right) + \left( \bar\chi_+, \phi \psi_- \right)   + h \left[ \left( \psi_-^T , \phi \gamma_2 \chi_+ \right) - \left( \bar\chi_+, \gamma_2 \phi^* \bar\psi_-^T \right) \right] \right\} \nonumber~.
\end{eqnarray}
Here $\phi_x = e^{i \eta_x}$ is the unitary higgs field and we do not show its kinetic term as it is the same as in (\ref{totalL}). The brackets  indicate both summation over coordinates and an inner product of spinors, for example   $\left( \bar\psi_- , \phi^* \chi_+ \right) \equiv \sum\limits_x  \bar\psi_{- \; x}  \phi^*_x \chi_{+\; x}$ and $ \left( \bar\psi_+, D_1 \psi_+\right) \equiv \sum\limits_{x,y}  \bar\psi_{+ \; x} D_{1 \; x y} \psi_{+
y}$.  There are  two Yukawa couplings  in the model,  $y$ and $y h$, which are both taken real. The coupling $h$ measures the ratio of the  Majorana to Dirac mass, while $y$ is the overall strength of the Yukawa coupling. The $S_{mirror}$ term above is the analogue of (\ref{DMmass}) in the ``345" theory.

When $y = h = 0$, the lattice action (\ref{toymodel}) has four global $U(1)$ symmetries, as every chiral component can be rotated independently, as  in Section 2. When both $y$ and $h$ are nonzero, there are only two $U(1)$ symmetries, acting on $\psi_+$ and $\chi_-$, respectively. The first is the anomalous global part of the gauge group and the second is the global symmetry of the spectator fermion.  When $h = 0$, the Majorana mass vanishes, and  we have one extra exact $U(1)$ that also acts on the charged fields and leads to extra zero modes. Hence,  we also need $h \ne 0$, as we will 
explicitly see below; note that the phase diagram for a four-dimensional model very similar to ours, without Majorana mass terms has recently been considered in \cite{Gerhold:2006rc}, using analytical methods.\footnote{We also need $h \ne 1$:  at $h=1$
the mirror fermion spectrum has an exact zero mode at finite $N$ for
arbitrary $\phi_x$ backgrounds, see Section \ref{splitofZ}.} 
  
  The continuum target theory that (\ref{toymodel}) is conjectured to flow to is the chiral Schwinger model, an anomalous $U(1)$ gauge theory with a single right-handed massless fermion of unit charge ($\psi_+$), a model solved  in  \cite{Jackiw:1984zi, Halliday:1985tg}. 
Since our paper is confined to the study of only the mirror fermion-Higgs dynamics,  without gauge fields and without taking  massless fermion loops into account, we do not have to  cancel  anomalies of the light spectrum;  minimality of the field content is why the ``toy" model was chosen as an arena to study the strong mirror-sector dynamics.

The fermion action (\ref{toymodel}),  the gauge action,  and  $S_{\kappa}$ (the action of $\phi$ in (\ref{Skappa})), combined with the usual Dirac fermion measure and the measures for the $U(1)$ link fields and   $\phi_x$,
completely define the lattice partition function of the toy model. 
As a matter of principle, the lattice model defined above could be, right away, subjected to numerical simulations including both dynamical gauge fields and fermions.
However, this is beyond the scope of  this paper,  which is restricted to  the $g=0$ case. Clearly, this is  only a first relatively simple step in the full analysis of the dynamics and the utility of the proposal.  Working out  the dynamics at $g=0$  is also  a litmus test for the proposal: if the analysis fails to yield a chiral spectrum or an unbroken gauge symmetry (more precisely---its global counterpart) at this point, the proposal has to be abandoned and no further  study at $g\ne 0$ is warranted.

\subsection{The toy model partition function at $g=0$}
\label{splitofZ}

 When $g \rightarrow 0$, only zero-action fluctuations of the gauge field  contribute to the path integral. In other words, $U_{x,y} = \omega_x \;\omega^\dagger_y$ with $\omega_x = e^{i \alpha_x}$. The $\omega_x$ fields   appear in the kinetic terms of (\ref{toymodel}) and in the Yukawa couplings, via the projectors used to define chirality components $\bar\psi_\pm$ of the charged  $\bar\psi$. We can remove the $\omega_x$ dependence in the lagrangian by a local field
  redefinition of  $\bar\psi$, $\psi$, and $\phi$:
\begin{equation}
\label{psiredefin}
\psi_x = e^{i \alpha_x} \; \psi_x^\prime, ~~ \bar\psi_x = \bar\psi_x^\prime \; e^{-i \alpha_x}~, ~~ \phi_x = e^{-i \alpha_x} \phi^\prime_x~, 
\end{equation}
with unit Jacobian. Thus the path integral over the $\omega_x$ decouples, yielding the volume of the group. We are left,   dropping the primes, with a nontrivial  integral over $\psi$, $\chi$, and $\phi$.

Now, as explained in Section 2, the partition function in a trivial gauge background splits into a product of mirror and light partition functions. As also explained there, for the purpose of  the study of the $g=0$ dynamics of the mirror fermion sector, the chiral split of the measure should be worked out explicitly, in order to    make the decoupling of the light and mirror sector  manifest and to facilitate a Monte  Carlo study of the mirror sector dynamics.

To work  out the   $\pm$-chirality split, it is useful  to define the measure in terms of  eigenvectors of the Neuberger operator. Since the chiral projectors involve the Neuberger operator, using its eigenvectors to define the measure facilitates an easy split of the measure into chiral components. The details are straightforward but somewhat tedious and are given  in Appendix A. For the reader interested in the  final formulae, the important expressions are eqn.~(\ref{psifinal})---the expansion of the GW-chirality components of spinors in terms of GW momentum eigenstates---and eqn.~(\ref{measure}), the final form of the measure relevant for its chiral split (also reproduced here in (\ref{lightmeasure}), (\ref{mirrormeasure})).
 In this section, we only present the  final expression for the partition function and measure resulting from applying these formulae. 

We start with  the  partition function written in terms of  position space variables, the charged
$\left\{ \psi_{\pm \; x}, \bar\psi_{\pm \; x} \right\} $ and neutral $\left\{ \chi_{\pm \; x},  \bar\chi_{\pm \; x}\right\} $, in terms of which the chiral split of the measure is quite complicated. 
Then, we change variables, using  (\ref{psifinal}), in both the action and measure 
in terms of appropriately chosen---corresponding to the GW eigenvalues with positive imaginary part---momentum space variables $\left\{ \psi_{\pm \; x}, \bar\psi_{\pm \; x} \right\}  \rightarrow \left\{ \alpha_{{\bf k}\; \pm}, \bar\alpha_{{\bf k}\; \pm} \right\}$ and
$\left\{ \chi_{\pm \; x}, \bar\chi_{\pm \; x} \right\} \rightarrow \left\{ \beta_{{\bf k}\; \pm}, \bar\beta_{{\bf k}\; \pm} \right\}$.  After performing all the required summations over $x,y$,  
the results for  the various components of $S_{light}$ and $S_{mirror}$ are as described below.

The kinetic terms for the mirror and light fermions  from (\ref{toymodel}) become, see (\ref{GWactionalpha}):
\begin{equation}
\label{mirrorkinetic}
S_{mirror}^{kin} = \sum\limits_{\bf{k}} \; \lambda_{\bf{k}} \;  \left( \bar\alpha_{\bf{k} -} \alpha_{\bf{k} -} + \bar\beta_{\bf{k} +} \beta_{\bf{k} +}  \right) ~, 
\end{equation}
\begin{equation}
\label{lightkinetic}
S_{light}^{kin} = \sum\limits_{\bf{k}} \; \lambda_{\bf{k}} \;  \left( \bar\alpha_{\bf{k} +} \alpha_{\bf{k} +} + \bar\beta_{\bf{k} -} \beta_{\bf{k} -}  \right) ~.
\end{equation}
Here $\lambda_{\bf{k}}$ is the eigenvalue of the free Neuberger
operator with positive imaginary part,  eqn.~(\ref{eigenvaluesGW}),
corresponding to momentum $\bf k$. The most important properties of 
$\lambda_{\bf{k}}$ are that $\lambda_{\bf{k}}-1$ is a complex number
of unit modulus, a consequence of the GW condition, and 
that $\lambda_{\bf k}$ only has a zero at $\bf{k} = 0$, see (\ref{eigenvaluesGW}). The momentum sums here and below are over $k_1, k_2 = 1, ..., N$.

The measure of the $g=0$ partition function, see eqn.~(\ref{measure}),  splits naturally into  a measure for the light fermions $\alpha_{\bf{k} \; +}, \bar\alpha_{\bf{k} \; +}, \beta_{\bf{k} \; -}, \bar\beta_{\bf{k} \; -}$: 
\begin{equation}
\label{lightmeasure}
\prod\limits_{x} d \bar\psi_{+ \; x} d \psi_{+ \; x} d \bar\chi_{-\; x } d \chi_{-\; x} \equiv  \prod\limits_{k_1, k_2 = 1}^{N} 16 (1 - \lambda_{\bf{k}}^*) d \alpha_{\bf{k} +} d\bar\alpha_{\bf{k}+} d \beta_{\bf{k} -} d\bar\beta_{\bf{k}-} ~,
\end{equation}
and a measure for the mirror fermions $\alpha_{\bf{k} \; -}, \bar\alpha_{\bf{k} \; -}, \beta_{\bf{k} \;+},\bar\beta_{\bf{k} \;+}$:
\begin{equation}
\label{mirrormeasure}
\prod\limits_{x} d \bar\psi_{- \; x} d \psi_{- \; x} d \bar\chi_{+\; x } d \chi_{+\; x} \equiv  \prod\limits_{k_1, k_2 = 1}^{N} 16 (1 - \lambda_{\bf{k}}^*) d \alpha_{\bf{k} -} d\bar\alpha_{\bf{k}-} d \beta_{\bf{k} +} d\bar\beta_{\bf{k}+} ~.
\end{equation}

To complete  the definition of the mirror fermion partition function $Z_{mirror}$, we need to write the Yukawa couplings of the mirror fermions in $S_{mirror}$ (\ref{toymodel}) to the unitary Higgs field. To this end, 
we first define the Fourier transforms of $\phi_x = e^{i \eta_x}$, with $\omega_N \equiv e^{\frac{2 \pi i}{N}}$:
\begin{eqnarray}
\label{phimomentum}
\Phi_{\bf{k}} \equiv \frac{1}{N^2} \sum_{\bf{x}} \omega_N^{- \bf{k} \cdot \bf{x}} \; e^{i \eta_x}~~,~~
\Phi^*_{\bf{k}} \equiv \frac{1}{ N^2} \sum_{\bf{x}} \omega_N^{- \bf{k} \cdot \bf{x}}\;  e^{- i \eta_x} = (\Phi_{- {\bf{k}}})^* ~.
\end{eqnarray}
 The two Yukawa couplings in the toy model lagrangian are, then: 
\begin{eqnarray}
\label{diracmirror}
 {1 \over y}S_{mirror}^{Dirac} &=& \frac{1}{2} \sum\limits_{\bf{k}, \bf{p}}  \;(2 - \lambda_{\bf{k}}) \;
\left( \bar\alpha_{\bf{k}-} \beta_{\bf{p}+}\Phi^*_{\bf{k}-\bf{p}} + \bar\beta_{\bf{k}+} \alpha_{\bf{p}-} \Phi_{\bf{k}- \bf{p}} e^{i (\varphi_{\bf{p}} - \varphi_{\bf{k}})}  \right), \nonumber \\
{1 \over y h}S_{mirror}^{Maj}  &=&
\end{eqnarray}
\begin{eqnarray}
i\sum\limits_{\bf{k}, \bf{p}}  
 \left[ \alpha_{\bf{k}-} \beta_{\bf{p}+}\Phi_{-\bf{k}-\bf{p}}e^{i \varphi_{\bf{k}}}  -
\bar\beta_{\bf{k}+} \bar\alpha_{\bf{p}-} \Phi^*_{\bf{k}+ \bf{p}} 
\; {(2 -  \lambda_{\bf{p}})(2- \lambda_{\bf{k}}) e^{- i \varphi_{\bf{k}} }  - \lambda_{\bf{p}} \lambda_{\bf{k}} e^{- i \varphi_{\bf{p}} } \over 4}  \right],\nonumber
\end{eqnarray}
where the momentum-dependent phase factors $e^{i \varphi_{\bf k}}$ are defined in eqn.~(\ref{phi}).

The existence of an exact mirror-fermion zero mode at $h=1$, alluded to in Section \ref{deftoy} and evident in all our numerical results, for arbitrary $\phi_x$ backgrounds, can now be shown. The  mirror zero mode at $h=1$ is a Majorana-Weyl fermion: this can be seen   by  letting all but the ${\bf{k}}=0$ components of $\beta_{\bf{k} +}, \bar\beta_{\bf{k} +}$ vanish,  taking  $\beta_{{\bf 0} +} = - i \bar\beta_{{\bf 0} +}$, and showing that the mirror action (\ref{diracmirror}) vanishes if $h=1$.

The equations presented in  this section completely define the  partition function. The decoupling of the light and mirror partition functions is   manifest. 
The mirror partition function $Z_{mirror}$ is defined by means of the  fermion measure (\ref{mirrormeasure}), the  measure for the unitary  Higgs field, $\prod\limits_{x}  \int\limits_0^{2\pi} d \eta_x$,  and the action $S_{mirror} = S^{kin}_{mirror} + S^{Dir}_{mirror} + S^{Maj}_{mirror} + S_\kappa$, with the various terms defined in (\ref{mirrorkinetic}), (\ref{diracmirror}), and (\ref{Skappa}). 

As mentioned in the Introduction, the split of the measures allows us, from now on, to concentrate on  studying the properties of the mirror fermion-Higgs sector with partition function $Z_{mirror}$. 

\subsection{Evidence for the existence of a symmetric phase  at strong Yukawa coupling}
\label{strongsymmetric}

In this Section, we present our study of the symmetry 
realization in the mirror theory at strong Yukawa coupling.
We remind the reader that the two possibilities are:
\ben
\item
A disordered (``symmetric'') phase, where correlation lengths
are of order the lattice spacing.  The absence
of order in the Higgs field is the analogue
of the symmetric phase in four dimensions.
The gauge boson would remain massless,
apart from the mass generated by the Schwinger mechanism.
\item
A quasi-ordered phase, where correlations
fall off according to a power law.  In this
phase the gauge boson would acquire a mass-squared
of order $\kappa g^2$.  It
is the two-dimensional analogue of spontaneous
gauge symmetry breaking in four dimensions.
\een
In the quasi-ordered phase there are long wavelength
modes that could mix with fermion bilinear composites
and form light mirror sector states.  Hence any
indication of critical behavior in susceptibility
data will be a sign of possible problems for the
decoupling of the mirror sector.
To this end, we perform  a Monte Carlo simulation of 
the mirror theory, whose partition function was  
defined in Section \ref{splitofZ}; all averages are 
computed using the Monte Carlo techniques described in Appendix \ref{simdetails}.  

Apart from a few remarks below, all simulations are done in 
the limit when the kinetic term of the mirror fermions, 
eqn.~(\ref{mirrorkinetic}), is neglected, i.e. 
in the strong Yukawa coupling limit $y\rightarrow \infty$
(cf. Appendix \ref{finitey}).
The ratio of Majorana to Dirac Yukawa couplings, the parameter $h$  in (\ref{diracmirror}), is kept fixed.  For $N=4,8,16$, our results indicate that  in this limit: 
\bit
\item
There is no sign problem for the fermion determinant at $h>1$. We have no analytic proof for arbitrary $\phi_x$ background (we have been able to show positivity for a constant  background and to first order in small   fluctuations). However, based on our simulations of the determinant in many (tens of millions) arbitrary $\phi$ backgrounds, not a single configuration has lead to a negative determinant for $h>1$. The results of the simulations and the presence of the exact zero mode at $h=1$  suggest that it should be possible to find an analytic  proof of positivity.
\item
For $0<h<1$ there is a sign problem and indications of critical behavior, as described below.
\item
All correlation lengths, for $\kappa = 0.1, 0.5 $  (both smaller than the $XY$-model $\kappa_c \simeq 1.12$)  are $\ord{a}$ and there is  no critical behavior for $h>1$ as discussed below.
\item
Within errors, for all $h>0$, there is no direction of $\eta$ that
is favored.  The shift symmetry $\eta \to \eta + {\rm const.}$ seems
to be preserved.  However, this is not a test of quasi-order, as  we need to look at the
response to a small external perturbation---hence, we perform the susceptibility measurements.
\item
The susceptibility, for the same values of $\kappa$, shows a large response for $h<1$
that increases with volume,
indicative of quasi-order, and
a small response for $h>1$, indicative of the
absence of quasi-ordering.
\item The vortex density decreases as $h$ decreases towards $h=1$, as in the Berezinski-Kosterlitz-Thouless transition towards the low-temperature ``confined-vortex" phase.
\eit

In this summary, and in the discussion that follows, it
is important to keep in mind that the phase transition
at $h=1$ is not sharp at finite $N$.  Therefore when
we speak of properties at, say, $h>1$, what is implied
is behavior at $h>1+\e$ with $\e>0$ a small number
that tends to zero as $N\to\infty$.

\subsubsection{Scalar field susceptibility}  
\label{scalarsusc}

We begin with a study of the scalar field susceptibility, 
which is a measure of the long distance correlations of the $XY$-model field $\phi_x = e^{i \eta_x}$.  In the "high-temperature" disordered phase of the $XY$ model the susceptibility is small, of order the lattice spacing or less, depending on the value of $\kappa < \kappa_c$. The critical coupling for the pure $XY$ model on a square lattice is known to be $\kappa_c \sim 1.12$ \cite{Wolff:1988uh}. 
In the "low-temperature" phase with algebraic long-range order, the susceptibility increases with the volume of the system. 

The susceptibility, which can also be thought of as the zero-momentum propagator of $\phi_x$, is defined in the usual way:
\beq
\label{chiscalar}
\chi = \frac{1}{N^2} \bigvev{ \left| \sum_x \; e^{ i \eta_x} \; \right|^2 }~.
\eeq
For an explanation of the meaning and numerical implementation of the $\langle \ldots \rangle$ average, see Appendix \ref{simdetails}.

We now present the results of numerical simulations of the scalar
field susceptibility (\ref{chiscalar}). 
First, in Fig.~\ref{fig:kappa0.1scalar}, we show the 
susceptibility for $\kappa  = 0.1$ for $N=4,8,16$, as a 
function of $h$; the pure $XY$-model values are shown by
horizontal dashed lines. We see that for $h>1$ the fermions do not have an appreciable effect on the scalar field susceptibility. 
Similar results are obtained for $\kappa = 0.5$, with the resulting overall rise in susceptibility, due to the closeness of $\kappa$ to $\kappa_c  = 1.12$, and are shown on Fig.~\ref{fig:kappa0.5scalar}. 
Finally, in Fig.~\ref{fig:kappazero}, we give the result 
for $\kappa = 0$.  It is qualitatively similar to the
$\kappa=0.1, 0.5$ results.

The results of the simulations of the scalar susceptibility 
displayed here show that for $h>1$ the fermions have a negligible 
effect on the scalar field susceptibility. 
The results of this Section support the conjecture 
that at strong Yukawa coupling the theory remains in the 
symmetric phase. We also note that the susceptibility 
of the theory with fermions, for $h>1$, is somewhat 
lower than the susceptibility of the pure $XY$-model with 
the same value of $\kappa$, for $N \gg 1$. 
This can be interpreted as a small 
renormalization of $\kappa$ to smaller values due to the fermions, 
which  appear to push  the theory further into the symmetric phase. 

On the other hand, for $h \leq 1$, there are indications of critical behavior and the scalar susceptibility grows with the volume of the system, as appropriate for the low-temperature phase of the $XY$ model. Also, for $h \simeq 1$, the fermion determinant vanishes and  there is a massless mirror fermion state (see the footnote in Section \ref{deftoy} and the numerical analysis of the spectrum in Section \ref{chiralmassless}). Clearly, at these values of $h$, the sensitivity to $y < \infty$ is enhanced, see Section \ref{finitey} in the Appendix \ref{simdetails}. 

Finally, for $h \rightarrow 0$, in Table \ref{tab1} we show the $\kappa=0.1$ results
for the Higgs susceptibility.  Clearly there is
a finite $h \to 0$, $N \to \infty$ limit, contrary
to the finite-size scaling that would occur in a
quasi-ordered phase.  We conclude that the
Higgs is in the disordered phase also for $h=0$, in agreement with
the analytic \cite{Gerhold:2006rc} and numerical   \cite{GJ2}  results in a similar model ($h=0$)  in four dimensions.
\begin{table}
\begin{center}
\begin{tabular}{|l||l|l||l|l||l|l||} \hline
$h$   & $N$ & $\chi$    & $N$ & $\chi$   & $N$  & $\chi$ \\ \hline
0.01  &  4  & 3.15(7) &  8  & 3.42(16) & 16   & 3.42(14) \\
0.005 &  4  & 3.15(6) &  8  & 3.41(16) & 16   & 3.41(14) \\
0.001 &  4  & 3.14(7) &  8  & 3.41(16) & 16   & 3.40(14) \\ \hline
\end{tabular}
\caption{\small $\kappa=0.1$ results for small values $h\rightarrow 0$; note that no $h \to 0$ extrapolation is
required, as the limit is already obtained within error and the  susceptibility  $\chi$ appears to approach a constant value as $N \to \infty$.
\label{tab1}}
\end{center}
\end{table}

\subsubsection{Binder cumulant}
\label{bindersection}

The Binder cumulant \cite{Binder:1981sa} is a quantity probing higher-order correlations, defined  as:
\beq
U = 2 - \frac{\vev{|M|^4}}{\vev{|M|^2}^2}, 
\label{udf}
\eeq
where $M$ is the total ``magnetization:"
$
M = \sum_x \phi_x, ~~
|M|^2 \equiv M M^*, ~~  |M|^4 \equiv (|M|^2)^2
$. The definition \myref{udf} is chosen such that in the pure $XY$-model $U$ interpolates between $0$ and $1$ as   the temperature decreases from infinity to zero: 
\beq
\lim_{\kappa \to 0} U = 0, \quad
\lim_{\kappa \to \infty} U = 1~.
\label{uli}
\eeq
This is easily seen by noting that for  $\kappa=0$, there is no intersite correlation and hence $U$ can be computed by treating the $\phi_x$ as random, delta-correlated, field leading to $U=0$. On the other hand, 
for $\kappa=\infty$, there is perfect order in each configuration and the $\phi_x$ are frozen, giving rise to $U=1$. 

We show the results of the measurement of the Binder cumulant for two values of $\kappa$:  $\kappa = 0.1$ in Fig.~\ref{fig:kappa0.1binder} and   $\kappa = 0.5$ in Fig.~\ref{fig:kappa0.5binder}. In each case, as a function of $h$ the Binder cumulant behaves as if $h \leq 1$ corresponds to the broken phase of the $XY$ model and $h > 1$---to the unbroken phase, consistent with the measurement of the scalar susceptibilities of the previous section, and thus providing further evidence for the persistence of the symmetric phase at strong Yukawa coupling and $h>1$.

\subsubsection{Vortex density}
\label{vortexsection}

The Berezinski-Kosterlitz-Thouless transition in the $XY$-model can be interpreted, 
see \cite{id}, as due to  the deconfinement of vortices above the critical temperature (at $\kappa \le \kappa_c$).  The definition and algorithm of finding the vortex density---the number of vortex-antivortex pairs per $XY$ spin---is described in \cite{tobochnik}. 

We studied the vortex density as a function of $h$.
We show the results,  for $\kappa=0.5$, on Fig.~\ref{fig:kappa0.5vortex}. It is clear from the figure that the vortex density  of the deformed $XY$ model at $h>1$ is slightly higher than the value appropriate to the pure-$XY$ model for $\kappa=0.5$; this is consistent with the  susceptibility measurements, where the small effect of the fermions at large $y$ is to push the theory further into the symmetric phase.

{\flushleft{T}o} summarize, in Sections \ref{scalarsusc}, \ref{bindersection}, and \ref{vortexsection},  we presented measurements of the pure scalar probes of symmetry breaking---the scalar susceptibility, the Binder cumulant, and the vortex density. The results give strong support in favor of 
the existence of 
a symmetric phase at strong Yukawa coupling, at $h>1$, $\kappa < 1$.

\subsubsection{Fermion bilinear susceptibilities}
\label{fermionsusc}

In this Section, we consider another probe of symmetry breaking---the fermion susceptibilities. 
As discussed in the Introduction, a    scalar $\phi_x$ and a charged mirror fermion, $\psi_-$,  can produce a 
neutral fermion bound state at strong coupling. Similarly, one expects that  a charged and a neutral fermion,  $\psi_-$ and $\chi_+$, can bind into scalar bound states with the quantum numbers of $\phi_x$. 

The two composite charged complex scalars of lowest dimension that we can construct out of bilinears of the mirror fermions are $  \bar\psi_{-  } \chi_{+ } $ and $\psi_{-  }^T \gamma_2 \chi_{+ }$.  We then define the corresponding susceptibilities;  similar to the scalar case (\ref{chiscalar}) they can be interpreted as the zero-momentum propagators of the corresponding composite scalar fields.
Thus, we introduce the ``Dirac" susceptibility: 
\begin{eqnarray}
\label{diracsusc}
\chi_F &\equiv& \sum_x  \langle \bar\psi_{- \; x} \chi_{+ \; x} \; 
\bar\chi_{+ \; y} \psi_{- \; y} \rangle ~,
 \end{eqnarray}
  and the 
  ``Majorana" susceptibility:
\begin{eqnarray}
\label{majoranasuscept}
\chi_F^\prime &\equiv& \sum_x  \langle \psi_{- \; x}^T \gamma_2 \chi_{+ \; x} \; \bar\chi_{+ \; y} \gamma_2 \bar\psi_{- \; y}^T \rangle  ~. \end{eqnarray}
In four dimensions a disconnected part would also have to be subtracted; however, 
the disconnected part always vanishes in two dimensions due to the
absence of spontaneous symmetry breaking. 
The explicit form of $\chi_F$ and $\chi_F^\prime$ in 
terms of averages over the variables of integration 
introduced in Section \ref{splitofZ} is given in eqns.~({\ref{fermionsuscept}, \ref{fermionsuscept2}) of  Appendix \ref{fermionGFappx}. Note that the dimensionless fermion susceptibilities scale as $Y^{-2}$, at large dimensionless $Y\equiv ya$. 

The results for the modulus and argument of the 
Dirac susceptibilities are given, for $\kappa = 0.5$
in Figs.~\ref{fig:chiF1} and \ref{fig:chiF2}, respectively. 
Similarly, for the Majorana susceptibility, the results 
for $\kappa = 0.5$ are shown in Figs.~\ref{fig:chiFp1} 
and \ref{fig:chiFp2}, respectively. The same plots of the magnitude and arguments of the Dirac and Majorana fermion susceptibilities, this time for the smaller value of $\kappa = 0.1$ are given in Figs.~\ref{fig:chiF10.1}, \ref{fig:chiF20.1}, \ref{fig:chiFp10.1}, \ref{fig:chiFp20.1} and lead to the same qualitative conclusions.

 We observe that 
the behavior of their modulus follows, as a function of $h$, 
the behavior of the scalar susceptibility and the Binder cumulant.
We note the following dependence on system volume:
\bit
\item
For $\kappa=0.1$ and $h>1$, the fermion susceptibilities never
show a trend of increasing with $N$.  While the curves for $\kappa=0.5$ are not monotonic, there we also find 
no evidence for   finite-size scaling that would indicate the
presence of light modes.
\item
$\chi'_F$ shows a trend of increasing for $h<1$, for either
$\kappa$.  For $\chi_F$, the trend depends on $\kappa$.
This indicates that the two susceptibilities
are indeed independent probes of long distance
behavior.  The rise in either with $N$ is indicative
of light modes that exist at $h<1$.
\eit
The numerical results for the phases of the Dirac   and Majorana susceptibilities  at $h>1$ are consistent with phases equal to $0$ and $\pi$, respectively.

As a matter of comparison, we  have also studied the fermion susceptibilities in the broken phase, upon   increasing $\kappa > \kappa_c$. In the broken phase, the fermion susceptibilities (as well as the scalar susceptibility) show a dramatic increase with the volume of the system, which is clearly absent at $\kappa < 1$.

{\flushleft O}verall, our results for the fermion susceptibilities show no sign of critical behavior at $h>1$, and thus provide more evidence for the existence of a strong Yukawa symmetric phase. 

\subsection{Evidence for the absence of massless mirror 
	fermions at strong Yukawa coupling}
\label{chiralmassless}
In this Section, we turn to the fermion spectrum at 
strong Yukawa coupling $y \gg 1$ and $\kappa < 1$.  
We study two kinds of correlators:  those that probe 
the neutral fermion spectrum and those probing the 
charged fermion sector.  All correlators that violate 
the $U(1)$ symmetry, i.e.~of the form $\bar\psi_+ \chi_-$, 
vanish, due to the absence of spontaneous
symmetry breaking of a continuous symmetry
in two dimensions\footnote{As a check on
our numerical simulations, we have verified
this property in both the disordered and quasi-ordered
phases.} \cite{MWC}.

In the disordered phase the mirror fermions obtain 
mass by the mechanism explained in the 
Introduction \cite{Witten:1978qu}, 
\cite{Eichten:1985ft, Hasenfratz:1988vc, Stephanov:1990pc, Golterman:1990zu,Golterman:1991re}: 
a charged fermion can bind with the charged scalar 
and pair up with the neutral fermion into a heavy 
neutral bound state.  It is also clear that in the 
disordered phase, the charged mirror fermion spectrum 
(if charged stable single-particle states exist)  
is vector-like:  the neutral mirror fermion can bind 
with the scalar into a charged right-handed fermion, 
which then can pair up with the charged left-handed 
fermion into a heavy charged state.  We will study
correlators that probe the spectrum of both types.

Studying the fermion propagators in the $y \rightarrow \infty$  
limit will be a reliable guide to the spectrum of the mirror 
fermion theory at $y \gg 1$, so long as there are no massless 
states that invalidate the $1/y$-expansion and make 
the $y< \infty$ corrections important.
Further comments on the $y < \infty$ limit 
are given in Appendix \ref{finitey};
especially important is the numerical
result $\ln \Pf {\cal M} = \ln \det M$ mentioned there.
Thus, the absence of massless states for $h> 1$ that 
we find ensures the self-consistency of the analysis.
To probe the fermion spectrum at large $y$, we thus study the 
Green functions which do not vanish in the large-$y$ 
limit (note that at infinite $y$, all $++$ and 
$--$ correlators vanish).

To study the fermion spectrum of our model, we begin by first considering the charged fermions. Using $\psi_-$, $\chi_+$, and $\phi$,  the three simplest unit-charge local fermion operators can be defined:
\begin{equation}
\label{chargedfermions}
\eta_{1\; x}^c = \psi_{- \; x}~,~~ \eta_{2 \; x}^c = \chi_{+  \; x}\phi^*_x~,~~ \eta_{3 \; x}^c = \bar\chi^T_{+ \; x} \phi^*_x~. 
\end{equation}
Negative charge  fermions can, similarly, be created by the local operators:
\begin{equation}
\label{chargedfermions2}
\bar\eta_{1\; x}^c = \bar\psi_{- \; x}~,~~ \bar\eta_{2 \; x}^c = \bar\chi_{+  \; x}\phi _x~,~~ \bar\eta_{3 \; x}^c =  \chi^T_{+ \; x} \phi_x~. 
\end{equation}
The choice of this operator basis is motivated by simplicity and by the fact that the corresponding propagator (\ref{cfpropagator}) can also be easily computed in the quasi-ordered phase, see eqn.~(\ref{constantphi}), and the change of the spectrum observed as a function of $\kappa$.

The propagator of the charged fermions is the connected two-point function:
\begin{equation}
\label{cfpropagator}
D_{i j}^c ({\bf{k}}) = \sum\limits_x \omega_N^{- {\bf{k}}(x-y)} \langle \; \eta_{i  \; x}^c \;  \bar\eta_{j \; y}^c \rangle~, ~~i,j = 1,2,3.
\end{equation} 
Diagonalizing $D_{ij}^c({\bf{k}})$ by a (bi-)unitary transformation gives rise to a diagonal matrix with entries, which we denote by $1/\mu_i({\bf{k}}) (i=1,2,3)$, where the $\mu_i$ are the eigenvalues of the inverse charged fermion propagator. A massless mode would yield  a zero of some of the eigenvalues $\mu_i$ for some $\bf{k}$. 

The study of the spectrum is  simplified at infinite $y$, when all $++$ and $--$ Green functions of the mirror theory vanish (there is an extra symmetry of the partition function at $y\rightarrow \infty$  forbidding such nonzero expectation values).  Then it suffices to consider a reduced propagator, instead of (\ref{cfpropagator}), which we call $S^c$, which we obtain by keeping only the nonvanishing entries of (\ref{cfpropagator}):
\begin{eqnarray}
\label{SfermionChB}
S^c_{y-x} = \left( \begin{array}{cc}  \langle \psi_{- \; y}  \; \chi_{+\;  x}^T    \phi_x  \rangle &
 \langle \chi_{+\;  y}   \phi_y ^*\; \bar\psi_{- \; x} \rangle \cr
  \langle  \psi_{- \; y} \; \bar\chi_{+\;  x}   \phi_x \rangle  & 
\langle \bar\chi_{+\;  y}^T     \phi_y^* \; \bar\psi_{- \;x}  \rangle 
 \end{array} \right)~.
\end{eqnarray}
Clearly, the propagator $S^c$ obeys Tr$D^{c \dagger} D^c = {\rm Tr}S^{c \dagger}S^c$ in the infinite-$y$ limit.
Our Monte-Carlo simulation calculates  the quantity     $\Omega^{(2)}({\bf{k}})$, defined through the last equality below:
\begin{equation}
\label{defomega}
{\rm Tr} \; S^{c \dagger} S^c ({\bf{k}}) = \sum\limits_{i=1}^3 {1 \over |\mu_i({\bf{k}})|^2 }\equiv \left({1\over \Omega^{(2)} ({\bf{k}})}\right)^2~,
\end{equation}
where the Fourier transform is defined, similar to (\ref{cfpropagator}), as $S^c({\bf{k}}) = \sum_x \omega_N^{- {\bf{k}} (y - x)} S^c_{y-x}$. Eqn.~(\ref{defomega}) 
shows that a massless particle will manifest itself as a zero of $\Omega^{(2)}$ at some  $\bf{k}$. To look for  light mirror modes, we will  simply plot the value  of ${\rm min}_{\bf{k}} \Omega^{(2)}(\bf{k})$ as a function of $h$ and the lattice size $N$, for $y = \infty$.

Next, we consider the operators creating   neutral mirror fermions: 
\begin{equation}
\label{neutralfermions}
\eta_{1\; x}^n = \chi_{+ \; x}~,~~ \eta_{2 \; x}^n = \bar\chi_{+  \; x}^T ~,~~ \eta_{3 \; x}^n = \phi_x \psi_{- \; x}  ~, ~~ \eta_{4\; x}^n = \phi^*_x \bar\psi_{-\; x}^T~,
\end{equation}
with their conjugates defined as in (\ref{chargedfermions2}):
\begin{equation}
\label{neutralfermions2}
\bar\eta_{1\; x}^n = \bar\chi_{+ \; x}~,~~ \bar\eta_{2 \; x}^n =  \chi_{+  \; x}^T ~,~~ \bar\eta_{3 \; x}^n = \phi^*_x \bar\psi_{- \; x}  ~, ~~ \bar\eta_{4\; x}^n = \phi_x  \psi_{-\; x}^T~,
\end{equation}
and neutral fermion 
  propagator:
\begin{equation}
\label{nfpropagator}
D_{i j}^n ({\bf{k}}) = \sum\limits_x \omega_N^{- {\bf{k}}(x-y)} \langle \; \eta_{i  \; x}^n \;  \bar\eta_{j \; y}^n \rangle~, ~~i,j = 1,2,3,4.
\end{equation} 
In the $y\rightarrow \infty$ limit  it suffices to consider the reduced neutral propagator,  $S^n$, keeping only the nonvanishing entries of (\ref{nfpropagator}) and combining them into a matrix similar to (\ref{SfermionChB}):
\begin{eqnarray}
\label{SfermionNB}
S^n_{y-x} = \left( \begin{array}{cc} \langle \chi_{+\; y} \; \phi_{ x} \psi_{-\; x}^T   \rangle& \langle \chi_{+\;  y}  \; \phi_x^*  \bar\psi_{- \; x}  \rangle  \cr  \langle \bar\chi^T_{+ \; y} \; \phi_x \psi_{- \; x}^T \rangle&
 \langle \bar\chi_{+\;  y}^T  \;   \phi_x^*  \bar\psi_{- \; x} \rangle  
 \end{array} \right)~,
\end{eqnarray}
 obeying Tr$D^{n \dagger} D^n = 2 {\rm Tr}S^{n \dagger}S^n$ in the infinite-$y$ limit.
We define the quantity $\Omega^{(1)} (\bf{k})$ by an equation identical to (\ref{defomega}), but with $S^c$ replaced with $S^n$; our plots seeking to establish the presence or absence of massless neutral modes will similarly show the value of ${\rm min}_{\bf{k}} \Omega^{(2)}(\bf{k})$ as a function of $h$ and the lattice size $N$, for $y = \infty$.
Explicit expressions giving $S^c$ and $S^n$ in terms of the integration variables of Section \ref{splitofZ} are given in (\ref{Sfermion2}, \ref{SfermionCh2}) of Appendix \ref{fermionGFappx}. We use the expressions given there to calculate the $\Omega^{(1),(2)}$ in our Monte-Carlo simulation.

In the broken phase, when $\phi_x \equiv 1$ to leading order in perturbation theory,
the form of $S^c$ (and $S^n$, which coincides with it in this limit) and the perturbative spectrum, valid  when  $\kappa \rightarrow \infty$, are worked out in Appendix \ref{fermionGFappx}. 
An important check on our Monte Carlo simulation is that, for $\kappa \gg \kappa_c$ it reproduces the perturbative spectrum of eqns.~(\ref{constantphiS3}, \ref{smalleigenvalues}), including all the small eigenvalues found analytically to leading order in perturbation theory, see eqn.~(\ref{smalleigenvalues}). Furthermore, the small deviations of the Monte Carlo results from the perturbative spectrum can be seen to scale as the expected perturbative corrections to the tree-level result arising from  spin-wave exchange,  $\sim h^2 \kappa^{-1}$ (we find that these corrections  become numerically significant for $h>2$ for the smallest eigenvalues of $s(\bf{k})^{-1}$, see (\ref{constantphiS3})).

The quantities ${\rm min}_{\bf k} \Omega^{(1)}_{\bf(k)}$ and ${\rm min}_{\bf k} \Omega^{(2)}_{\bf(k)}$ provide a lower bound on the smallest  eigenvalue of the neutral and charged inverse propagators, respectively. Finally, recall  the  large-$y$ scaling (and that $y$ has dimensions of mass)   $\Omega_{\bf{k}} \sim y$. Introducing a dimensionless Yukawa coupling $Y=ya$,  the  mass scales as $Y a^{-1}$. The continuum limit is  $N\rightarrow \infty$, $a \equiv L_{phys}/N$, with fixed system volume $L_{phys}^2$.

In Figure \ref{fig:om1_05}, we display  $N {\rm min}_{\bf{k}}\; \Omega_{\bf{k}}^{(1)}$---the lower bound on the minimal value of the neutral fermion eigenvalue---as a function of $h$ for $\kappa = 0.5$ and $N=4, 8, 16$. Figure~\ref{fig:om2_05}   displays the corresponding quantity for the charged fermions. 
  In Figures~\ref{fig:om1_01} and \ref{fig:om2_01}, respectively, we show the lower bound on the neutral and charged  eigenvalues for $\kappa = 0.1$.  In units of $L_{phys}^{-1}$, the lower bound on the mass  can be obtained 
 by multiplying the plotted value of   by the dimensionless  large Yukawa coupling $Y$.
  
The numerical results for the lower bounds on the neutral and charged fermion propagator eigenvalues, $\mu_i$,
 displayed in these figures,  imply that:
 \begin{itemize}
 \item The lower bounds on both the neutral and charged eigenvalues $\mu_i$, for large $Y$,   are $ \sim Y N^{\alpha}  \; L_{phys}^{-1}$, with a  $\kappa$-  and (more weakly) $h$-dependent exponent $\alpha > 0$. 
 This presents evidence that the mirror fermions decouple from the infrared physics in the $N\rightarrow \infty$, fixed $L_{phys}$  limit.
 \item It is clear, by comparing Figs.~\ref{fig:om1_05} and \ref{fig:om1_01}, that  the exponent $\alpha$ is an increasing function $\kappa$. To better study this dependence, we have also    measured 
 $\Omega_{\bf{k}}^{1,2}$, for $h=2$ and $\kappa = 0.6, 0.7, \ldots, 1.4$, in addition to values that we display. We found that in the interval $0.6 \le \kappa \le 0.8$ the exponent $\alpha \simeq 1$ (for larger  values of $\kappa$ appropaching $\kappa_c \ge 1$ the susceptibilities rise with $N$ and the spectrum begins to approach that of the broken phase, see comments below). Thus for $\kappa$ in this interval, the mirror fermion masses are cutoff scale times $Y$.
 \item If the lower bound on the fermion eigenvalues is actually saturated, one can conclude from the data that the charged fermions are generally heavier than the neutral ones. 
 \end{itemize}
 In summary, our results present evidence that there are no light states in either the neutral or the charged fermion mirror sector.

Finally, to compare with the fermion spectrum in the broken phase, the dashed lines on each figure show the 
minimum values of $\Omega$ for $N=4,8,16$, 
calculated to leading order in perturbation theory (strictly valid when 
$\kappa \rightarrow \infty$) in the 
broken phase.   This was done 
by using the broken phase propagator of 
eqn.~(\ref{constantphiS3}).  As already mentioned, 
these perturbative results are in excellent quantitative 
agreement with Monte Carlo simulations in the quasi-ordered phase (performed  for $\kappa = 10 \gg \kappa_c \simeq 1$).
The dips in the dashed lines correspond to the approximate   
solutions of (\ref{smalleigenvalues}), 
leading at finite $N$ to almost massless 
modes.\footnote{We caution against concluding from 
the plots that the broken phase massless modes 
indicated in (\ref{smalleigenvalues}) are not present 
in the continuum limit for some
values of $h$---it is easy to check that 
as $N$ becomes large, the dips completely 
overcome the $N$ enhancement from the scaling 
of $y$, as the eigenvalues become closely 
spaced on the unit circle.}
Our results, discussed above, show that in the symmetric phase  the corresponding 
massless modes are absent.

\bigskip

\bigskip

\noindent
{\bf \large Acknowledgements}

\vspace{5pt}
EP  thanks Bob Holdom for
useful discussions and   suggestions. JG  expresses appreciation
for extensive use of the FTPI computer cluster,
and technical help from Graham Allan.
JG was supported in part by the Department of
Energy grant DE-FG02-94ER40823 at the University
of Minnesota.  
EP is supported 
by the National Science and Engineering 
Research Council of Canada; he also gratefully acknowledges the hospitality of the Galileo Galilei Institute of Theoretical Physics in Florence during the initial stages of this work.

\myappendix
\mys{Expansion of spinors in terms of chiral GW eigenvectors and chiral split of the measure}
\label{Measuresplit}
\vspace{-0.1in}

In this Appendix, we derive an expansion of spinors in terms of Neuberger's operator eigenstates. This is useful in writing the chiral split of the measure in Section \ref{splitofZ}.
We denote the Neuberger operator by $D$, obeying $\{\gamma_5, D\} = D \gamma_5 D$ and $\gamma_5 D \gamma_5 = D^\dagger$.  Recall that the eigenvalues $\lambda$ of $D$ obey
$\lambda + \lambda^* = \lambda \lambda^*$ and thus lie on a unit circle in the complex plane 
centered at $1$. For every eigenvalue with Im $\lambda\ne 0$ there is another eigenvalue $\lambda^*$ whose eigenvector, a 2-component spinor, is $\Psi_{\lambda^*} = \gamma_5 \Psi_\lambda$. Here $D \Psi_\lambda = \lambda \Psi_\lambda$. The inner product notation we use is  $(\Psi_{\lambda^\prime}^\dagger, \Psi_\lambda) \equiv \sum_x \Psi_{\lambda^\prime \; x}^\dagger  \Psi_{\lambda \; x} = \delta_{\lambda^\prime, \lambda}$. Orthogonality of eigenvectors with different eigenvalues follows from hermiticity of $\gamma_5 D$. 
Chiral components for the unbarred fields $\Psi$ are defined in the usual way, 
$\Psi_\pm = P_\pm \Psi$, while  for the barred  $\bar\Psi$, $\bar\Psi_{\pm} = \bar\Psi \hat{P}_{\mp}$, where $\hat{P}_\pm = (1\pm \hat\gamma_5)/2$, with $\hat\gamma_5 = (1-D)\gamma_5$.

To define the Dirac fermion measure,  we 
first expand an arbitrary spinor configuration
in terms of the eigenvectors $\Psi_{\lambda \; x}$ 
of $D$ as follows:\footnote{Here and below, there is an implicit sum over
multiplicities associated with a given eigenvalue $\lambda$.}
\begin{equation}
\label{psiexpansion1}
\Psi_x = \sum\limits_{Im\lambda>0} \[ c_\lambda \; \Psi_{\lambda \; x}
+ c_{\lambda^*} \; \Psi_{\lambda^* \; x}~ \] ,
\end{equation}
where $c_\lambda$ are Grassmann numbers.\footnote{
Here, we have displayed the contribution of only the eigenvalues $\lambda \ne 0,2$;  the contribution of the eigenvectors of 
 the real eigenvalues $\lambda  = 0$ and $\lambda = 2$ are implicitly assumed  present and will be added later. The reason is that they are always (anti)chiral in the usual sense: for $\lambda = 0,2$  the GW chirality is equal, up to a sign, to the usual $\gamma_5$ chirality. In the free case of zero gauge background we will   explicitly find these eigenvectors and see that the $\lambda = 0$ ones (one $+$ and one $-$) corresponds to $\vec{k} = (N,N)$, while the $\lambda =2$ ones (one pair of $\pm$ each) are obtained  for the three values of momentum $\vec{k} = (N/2,N), (N,N/2), (N/2,N/2)$. }
The measure of the $\Psi$ integration is, therefore: 
\begin{equation}
\label{psimeasure}
\prod_x d \Psi_x = \prod\limits_{Im \lambda > 0} d c_\lambda d c_{\lambda^*}~.
\end{equation}
The contributions of the real eigenvalues $\lambda = 0,2$  are to be added both to (\ref{psiexpansion1}) and (\ref{psimeasure}). We note that there is no Jacobian in eqn.~(\ref{psimeasure}). This follows from the fact that the GW operator is normal ($\[ D^\dagger, D \] = 0$) and hence can be diagonalized by a unitary transformation; the Jacobians from $\Psi$ and $\bar\Psi$ transforms cancel, for any gauge field background. 

Our next goal is to rewrite the measure  (\ref{psimeasure}) in terms of chiral components. 
This is easily accomplished  for the unbarred field $\Psi_x$ (keeping in mind that $\Psi_{\lambda^*} = \gamma_5 \Psi_\lambda$). 
Explicitly, we rewrite (\ref{psiexpansion1}) as follows:
\begin{eqnarray}
\label{psiexpansion2}
\Psi_x &=& \sum\limits_{Im \lambda>0} \[ {c_\lambda + c_{\lambda^*} \over \sqrt{2} } {\Psi_{\lambda\; x}+ \Psi_{\lambda^*\; x} \over \sqrt{2} } + {c_\lambda - c_{\lambda^*} \over \sqrt{2}} {\Psi_{\lambda\; x} - \Psi_{\lambda^*\; x} \over \sqrt{2}} \] \nonumber \\
&=&   \sum\limits_{Im \lambda>0} \[ \alpha_{\lambda+} \Psi_{\lambda+\; x} +  \alpha_{\lambda - \; x} 
\Psi_{\lambda -} \] ~,
\end{eqnarray}
 where we defined:
 \begin{eqnarray}
 \label{psimeasure2}
 \alpha_{\lambda \pm} &\equiv& {c_\lambda \pm c_{\lambda^*} \over \sqrt{2}} \\
 \Psi_{\lambda\pm \; x} &\equiv& {\Psi_{\lambda \; x} \pm \Psi_{\lambda^*\; x} \over \sqrt{2}}~.
 \end{eqnarray}
Now we can use the relation (\ref{psimeasure2}) to express the measure (\ref{psimeasure}) in terms of the variables $\alpha_\pm$, with a Jacobian that is only a numerical factor:
\begin{equation}\label{psimeasure3}
 \prod_x d \Psi_x =  \prod\limits_{Im \lambda > 0} d c_\lambda d c_{\lambda^*} = \prod\limits_{Im \lambda > 0} 4\;   d\alpha_{\lambda -} d\alpha_{\lambda +}
\end{equation}
and the expansion (\ref{psiexpansion2}) to express the action in terms of the chiral integration variables $\alpha_\pm$. The zero and cutoff ($\lambda = 2$) modes are easily added to (\ref{psiexpansion2}), (\ref{psimeasure3}), as shown below.
 
Now consider the more interesting $\bar\Psi$ case. Note the following chain of relations:
$D \Psi_\lambda = \lambda \Psi_\lambda$ $\rightarrow$ $\Psi_\lambda^\dagger D^\dagger = \Psi_\lambda^\dagger \lambda^*$ $\rightarrow$ $\Psi_\lambda^\dagger \gamma_5 D \gamma_5 = \Psi_\lambda^\dagger \lambda^*$ $\rightarrow$ $\Psi_{\lambda^*}^\dagger D \gamma_5 = \Psi_\lambda^\dagger \lambda^*$ $\rightarrow$ $\lambda^* \Psi_{\lambda^*}^\dagger \gamma_5 = \lambda^* \Psi_{\lambda}^\dagger$, where the $\gamma_5$ hermiticity of $D$ was used. 
We continue by expanding the row spinor $\bar\Psi$ in terms of a complete set of $\Psi_\lambda^\dagger$ solutions as follows:
\begin{equation}
\label{psibarexpansion1}
\bar\Psi_x = \sum\limits_{Im \lambda>0} \[ \bar{c}_\lambda \Psi_{\lambda\; x}^\dagger + \bar{c}_{\lambda^*} \Psi_{\lambda^*\; x}^\dagger~ \],
\end{equation}
again omitting the contribution of the $\lambda = 0,2$ eigenvalues that should be always added in ones mind.  
Similar to (\ref{psimeasure}) we define the barred fermion measure: 
\begin{equation}
\label{psibarmeasure1}
\prod\limits_x d \bar\Psi_x \equiv \prod\limits_{Im \lambda > 0} d \bar{c}_\lambda d \bar{c}_{\lambda^*} 
\end{equation}
The set of relations above (\ref{psibarexpansion1}) are useful because they  allow us to rewrite (\ref{psibarexpansion1}) in terms of $\hat{P}_\pm$ components as follows:
\begin{eqnarray}
\label{psibarexpansion2}
\bar\Psi_{\pm \;x} \equiv ( \bar\Psi \hat{P}_{\mp} )_x
&=& \sum\limits_{Im \lambda>0} \[ \bar{c}_\lambda 
\; {\Psi_{\lambda \; x}^\dagger \mp \Psi_{\lambda^*\; x}^\dagger (1 - \lambda) \over 2 } 
+
 \bar{c}_{\lambda^*} \; { \Psi_{\lambda^*\; x}^\dagger \mp \Psi_{\lambda\; x}^\dagger (1 - {\lambda^*}) \over 2 }  \] \nonumber \\
 &=& \sum\limits_{Im \lambda>0}  {\bar{c}_\lambda \mp (1-\lambda^*) \bar{c}_{\lambda^*} \over \sqrt{2}} \; {\Psi_{\lambda\; x}^\dagger \mp \Psi_{\lambda^*\; x}^\dagger (1-\lambda) \over \sqrt{2}}  ~, 
\end{eqnarray}
where the first line was obtained by applying the projector to (\ref{psibarexpansion1}) and in the second we used $|1 - \lambda| = 1$. Thus, similar to the definition of $\alpha_\pm$ of (\ref{psimeasure2}), we can define:
\begin{eqnarray}
\label{psibarmeasure2}
\bar{\alpha}_{\lambda \pm} &=& {1 \over \sqrt{2}} \left( \bar{c}_\lambda \mp (1 - \lambda^*) \bar{c}_{\lambda^*}\right) \nonumber \\
\bar\Psi_{\lambda \pm\; x}&=&{1 \over \sqrt{2}} \left( \Psi_{\lambda\; x}^\dagger \mp \Psi_{\lambda^*\; x}^\dagger (1 - \lambda) \right)~,
\end{eqnarray}
We see from the above equation that the Jacobian of the transformation from $\bar{c}$ to $\bar\alpha_\pm$ variables depends on $\lambda$; explicitly, eqn.~(\ref{psibarmeasure1}) becomes:
\begin{equation}
\label{psibarmeasure3}
\prod\limits_x d \bar\Psi_x \equiv \prod\limits_{Im \lambda > 0} d \bar{c}_\lambda d \bar{c}_{\lambda^*} = \prod\limits_{Im \lambda > 0} 4 (1 - \lambda^*) d \bar{\alpha}_{\lambda -} d \bar{\alpha}_{\lambda +} ~.
\end{equation}

As a simple demonstration of the self-consistency of this procedure, we check that in the vectorlike case we can use the measure in the final form (\ref{psimeasure3}) and (\ref{psibarmeasure3}) to calculate the determinant of $D$. To this end, write first  the expansion of $\bar\Psi$ in terms of $\alpha_\pm$:
\begin{equation}
\label{psibarexpansion3}
\bar\Psi_x = \sum\limits_{Im \lambda > 0} {\bar\alpha_{\lambda_+} + \bar\alpha_{\lambda_-} \over \sqrt{2}} \; \Psi_{\lambda\; x}^\dagger  + {\bar\alpha_{\lambda_+} - \bar\alpha_{\lambda_-} \over \sqrt{2} (\lambda^* -1) } \; \Psi_{\lambda^*\; x}^\dagger~,
\end{equation}
(recalling that zero and cutoff modes should be added as well) also note that the $\lambda^*-1$ factor in the denominator does not vanish since $\lambda=1$ is not in the spectrum. Then, substitute (\ref{psibarexpansion3}) as well as the $\alpha_\pm$ expansion of $\Psi$ of eqn.~(\ref{psiexpansion2}) into the GW action to obtain, using the orthogonality of the $\Psi_\lambda$ wavefunctions and recalling that $\lambda = \lambda^*/(\lambda^*-1)$:
\begin{eqnarray}
\label{GWactionalpha}
(\bar\Psi, D \Psi) \equiv \sum\limits_{x} \bar\Psi_x (D\Psi)_x = \sum\limits_{Im \lambda > 0} \bar\alpha_{\lambda +} \alpha_{\lambda +} \lambda + \bar\alpha_{\lambda -} \alpha_{\lambda -} \lambda.
\end{eqnarray}
Now recall the $\bar \alpha_\pm$ measure
  (\ref{psibarmeasure3}) and the similar expression \myref{psimeasure3} (with constant Jacobian) for $\Psi$ to find that the path integral over the $\lambda \ne 0,2$ eigenvalues reduces to
  \begin{eqnarray}
  \label{GWdet1}
  \int \prod\limits_{x} d \Psi_x d\bar\Psi_x \; e^{(\bar\Psi, D \Psi)} &=& c \prod\limits_{Im \lambda > 0} \lambda^2 (1- \lambda^*) = c^\prime \prod\limits_{Im \lambda > 0} \lambda\lambda^* \nonumber \\
  &=&  c^\prime \prod\limits_{\lambda \ne 0,2} \lambda = c^\prime {\rm det}( D) \big\vert_{\lambda \ne 0,2}~,
  \end{eqnarray}
i.e. the determinant of $D$, as appropriate (the factor $c^\prime$ hides various numerical constants as well as the contributions of the $\lambda = 0,2$ eigenvalues). The point of this exercise was to show that the Jacobian factor 
in (\ref{psibarmeasure3}) 
was crucial to obtaining the right result.

We end this section by summarizing the important formulae that were derived: the 
expansions of the fermionic fields in terms of generalized chiral and antichiral Grassmann amplitudes
$\alpha_\pm, \bar\alpha_\pm$: 
\begin{equation}
\label{psi}
\Psi_x = \sum\limits_{Im \lambda > 0} \[
 {\alpha_{\lambda +} + \alpha_{\lambda -} \over \sqrt{2}}\;  \Psi_{\lambda \; x} + {\alpha_{\lambda +} - \alpha_{\lambda -} \over \sqrt{2}} \; \Psi_{\lambda^*\; x} \]
\end{equation}
\begin{equation}
\label{psibar}
\bar\Psi_x = \sum\limits_{Im \lambda > 0}
\[ {\bar\alpha_{\lambda +} +\bar\alpha_{\lambda -} \over \sqrt{2}} \; \Psi_{\lambda\; x}^\dagger  + {\bar\alpha_{\lambda +} - \bar\alpha_{\lambda -} \over \sqrt{2} (\lambda^* -1)} \; \Psi_{\lambda^*\; x}^\dagger  \]
\end{equation}
and an expression for the measure,  combining (\ref{psimeasure3}) and (\ref{psibarmeasure3}), in terms of these  Grassmann amplitudes:
\begin{equation}
\label{measure}
\prod\limits_{x} d \bar\Psi_x d \Psi_x \equiv c \prod\limits_{Im \lambda > 0} d\alpha_{\lambda +} d \alpha_{\lambda -} d \bar{\alpha}_{\lambda +} d \bar\alpha_{\lambda -} (1- \lambda^*)~.
\end{equation}
This defines a split of the measure in terms of chiral and antichiral fields. 
The zero and cutoff modes (the real eigenvalues of $D$) should be added to (\ref{psi}, \ref{psibar}, \ref{measure}). 

The usefulness of (\ref{psi}, \ref{psibar}) is that if we write the lagrangian in terms of $\Psi_\pm$, $\bar\Psi_\pm$ components, e.g., as in (\ref{toymodel}), we have a straightforward way to express the corresponding terms via the integration variables $\alpha_\pm$. For example, if a term contains $\bar\Psi_{+\;  x}$, we simply need use (\ref{psibar}) and put all $\bar\alpha_- $ to zero in that expansion, etc. 

\vspace{-0.1in}
\subsection{Explicit form for the  GW eigenvectors for $g = 0$.}

Here we work out the explicit form of the eigenvectors of the Neuberger-Dirac operator in the free case.
We begin by writing the free Neuberger-Dirac operator in momentum space:
\begin{eqnarray}
\label{fourierGW}
\Psi_x &=& {1 \over N} \sum\limits_{k_1, k_2 = 1}^N \omega_N^{\vec{k} \cdot \vec{x}} \; \tilde\Psi(k) ~, ~~~\omega_N \equiv e^\frac{2 \pi i}{N}~, 
\end{eqnarray}
\begin{eqnarray}
(D \Psi)_x &=&  {1 \over N} \sum\limits_{k_1, k_2 = 1}^N \omega_N^{\vec{k} \cdot \vec{x}} \; D({\bf{k}}) \; \tilde\Psi(k)~.
\end{eqnarray}
We also define $s_\mu \equiv   \sin ( \pi k_\mu/N)$ and   $c_\mu \equiv  \cos (\pi k_\mu/N)$ and the following functions of $k_\mu = 1, \ldots, N$, $\mu = 1,2$:
\begin{eqnarray}
\label{abc}
a_{\bf{k}} &\equiv&  1 - { 1 - 2 s_1^2   - 2 s_2 ^2  \over \sqrt{1 + 8 s_1^2 s_2^2 }      }~, \nonumber \\
b_{\bf{k}} &\equiv& {2 s_2 c_2 \over \sqrt{1 + 8 s_1^2 s_2^2 }      }~,\\
c_{\bf{k}}  &\equiv& {2 s_1 c_1 \over \sqrt{1 + 8 s_1^2 s_2^2 }      }\nonumber ~,
\end{eqnarray}
in terms of which $D(\bf{k})$ takes the explicit form (we use $\gamma_1 = \sigma_1, \gamma_2 = \sigma_2, \gamma_5 = \sigma_3$):
\begin{eqnarray}
\label{momentumGW}
D(\bf{k}) = \left( \begin{array}{c} a_{\bf{k}}   \cr i c_{\bf{k}} - b_{\bf{k}} \end{array} \begin{array}{c}   i c_{\bf{k}} + b_{\bf{k}} \cr  a_{\bf{k}} \end{array} \right)~.
\end{eqnarray}
The GW condition is equivalent to the relation $a_{\bf{k}}^2 + b_{\bf{k}}^2 + c_{\bf{k}}^2 = 2 a_{\bf{k}}$, which is easily verified from (\ref{abc}). The eigenvalues of $D(\bf{k})$ are: 
\begin{equation}
\label{eigenvaluesGW}
\lambda_{\bf{k}} = a_{\bf{k}} \pm i \sqrt{b_{\bf{k}}^2 + c_{\bf{k}}^2}~.
\end{equation}
With the help of this equation it is easily seen that the only real eigenvalues occur for ${\bf{k}} = (N,N), (N/2,N), (N, N/2)$ and $(N/2, N/2)$. The $(N,N)$ eigenvalue is the zero mode $\lambda = 0$, while the other three correspond to the cutoff value $\lambda = 2$. For every one of these eigenvalues there are two eigenvectors of opposite 
$\gamma_5$ chirality; these can be taken simply to be represented by $\left(\begin{array}{c}0\cr 1\end{array}\right)$ and $\left(\begin{array}{c}1\cr 0\end{array}\right)$, as will be done explicitly in eqn.~(\ref{cutofffreesolutions}).

The eigenvectors with Im$\lambda > 0$ are explicitly given by:
\begin{equation}
\label{normalizedeigenvectors}
\Psi_{\lambda_{\bf{k}} \;x} = {1 \over \sqrt{2} N } \; \omega_N^{ \vec{k} \cdot \vec{x}  } 
\:
\left( \begin{array}{c} 1 \cr { i b_{\bf{k}} + c _{\bf{k}} \over \sqrt{b_{\bf{k}}^2 + c_{\bf{k}}^2}  } \end{array} 
\right) ~,
\end{equation}
and are easily seen to obey:
\begin{equation}
\label{normalization}
\left( \Psi_{\lambda_{\bf{k}} }^\dagger, \Psi_{\lambda_{\bf{k}^\prime} }\right) 
\equiv \sum\limits_{\vec{x}} \Psi_{\lambda_{\bf{k}}\; x}^\dagger   \Psi_{\lambda_{\bf{k}^\prime\; x}}  = \delta_{\vec{k}, \vec{k}^\prime}~,
\end{equation}
a property assumed in deriving, e.g. (\ref{GWactionalpha}, \ref{GWdet1}). In particular, the expansions
(\ref{psi}, \ref{psibar}, \ref{measure}) hold by simply replacing the $\Psi_\lambda$ there by (\ref{normalizedeigenvectors}); note that the sums and products there are  over all $\vec{k}$ apart from the four points corresponding to real eigenvalues.

Now consider the  $\lambda = 2$ eigenvalues; the $\lambda = 0$ ones can be considered in exact parallel. We  proceed simply by recalling 
that $\lambda = 2$ eigenvalues are always (anti)chiral in the usual sense. More precisely, for the
 unbarred spinors, GW and usual chirality coincide for any $\lambda$, while for the barred spinors, 
GW and usual chirality are opposite for $\lambda = 2$; this  simply follows from the fact that for $\lambda = 2$ we have $\hat\gamma_5 = (1 - D) \gamma_5 = - \gamma_5$. In our basis of gamma matrices definite $\gamma_5$ chirality $\Psi_\pm$ spinors have the form:
\begin{eqnarray}
\label{cutoffchiralpsi}
\Psi_+ &=& \left( \begin{array}{c} a \cr 0\end{array} \right)~~, ~~
\Psi_- = \left( \begin{array}{c} 0 \cr b\end{array} \right)~~, ~~
\end{eqnarray}
while definite $\hat{\gamma}_5 = - \gamma_5$ chirality $\bar\Psi_{\pm}$ are $\bar\Psi_{\pm} \equiv \bar\Psi \hat{P}_{\mp} = \bar\Psi (1\mp \hat\gamma_5)/2\big\vert_{\lambda = 2} = \bar\Psi (1 \pm \gamma_5)/2$. Thus, the $\bar\Psi_\pm$ spinors with $\lambda = 2$ have the form:
\begin{eqnarray}
\label{cutoffchiralpsibar}
\bar\Psi_+ &=& \left( \; c \; \; 0 \;  \right)~~,~~ \bar\Psi_- = \left(\; 0 \; \; d \; \right)~.
\end{eqnarray}
 This allows us to write the $\lambda = 2$ solutions for two spinors $\Psi_- $ and $\Psi_+ $ (we use the fact that $\lambda=2$ solutions in the free case  occur for $\vec{k} = (N/2,N/2), (N, N/2), (N/2,N)$ only):
\begin{eqnarray}
\label{cutofffreesolutions}
\Psi_{- \; x} &=& {1 \over N}\left( \omega_N^{(N/2)x_1 + (N/2) x_2} \alpha_-  + \omega_N^{(N/2)x_1 + N x_2} \beta_-  + \omega_N^{Nx_1 + (N/2) x_2} \gamma_-  \right) \left( \begin{array}{c}0 \cr 1\end{array}\right)\\
\Psi_{+\; x}  &=& {1 \over N}\left( \omega_N^{(N/2)x_1 + (N/2) x_2} \alpha_+  + \omega_N^{(N/2)x_1 + N x_2} \beta_+  + \omega_N^{Nx_1 + (N/2) x_2} \gamma_+  \right) \left( \begin{array}{c}1 \cr 0\end{array}\right) \nonumber \\
\bar\Psi_{-\; x}  &=& {1 \over N}\left( \omega_N^{-(N/2)x_1 - (N/2) x_2} \bar\alpha_-  + \omega_N^{-(N/2)x_1 - N x_2} \bar\beta_-  + \omega_N^{-Nx_1 - (N/2) x_2} \bar\gamma_-  \right) \left( \; 0 \: \: 1\right) \nonumber \\
\bar\Psi_{+\; x}  &=& {1 \over N}\left( \omega_N^{-(N/2)x_1 - (N/2) x_2} \bar\alpha_+  + \omega_N^{-(N/2)x_1 - N x_2} \bar\beta_+  + \omega_N^{-Nx_1 - (N/2) x_2} \bar\gamma_+  \right) \left( \; 1 \; \; 0 \right) \nonumber~,
\end{eqnarray}
where the $\alpha, \beta, \gamma$ are the corresponding Grassmann integration variables of the $\lambda = 2$ modes; also clearly the various position factors are either unity or of the form $(-1)^{x_1}$, etc., which will be important when performing the summations after the further insertion in the action.

Now, we put all formulae together, using the momentum space eigenvectors of $D(\bf{k})$ found above, to obtain the final form of the expansion of $\Psi$, $\bar\Psi$ in terms of GW eigenvectors. To this end, we first define the phase factor from the eigenvectors (\ref{normalizedeigenvectors}), now extended to include all $\bf{k}$:
\begin{equation}
\label{phi}
e^{i \varphi_{\bf{k}}} \equiv \left\{ \begin{array}{cc} 
{i b_{\bf{k}} + c_{\bf{k}} \over \sqrt {b_{\bf{k}}^2 + c_{\bf{k}}^2 } } & {\rm if} \; {\bf{k}} \ne (N,N), ({N\over 2},N), (N, {N \over 2}), ({N\over 2},{N\over 2})  \cr
1  &\;  {\rm if} \;{\bf{k} }= (N,N), ({N\over 2},N), (N, {N \over 2}), ({N\over 2},{N\over 2})~\end{array} \right.~, 
\end{equation}
 and  introduce the unitary matrices $U_{\bf{k}}$ and $V_{\bf{k}}$:
\begin{equation}
\label{unitaryprojectors}
U_{\bf{k}} = {1\over 2} \;   \left( \begin{array}{cc}  2 - \lambda_{\bf{k}} &\lambda_{\bf{k}} 
e^{- i \varphi_{\bf{k}}} \cr - \lambda_{\bf{k}}^* e^{i \varphi_{\bf{k}} }& 2 - \lambda_{\bf{k}}^* \end{array} \right)        
 ~, ~~ V_{\bf{k}}   = {1\over 2} \;   \left( \begin{array}{cc}  (2 - \lambda_{\bf{k}}^*) e^{i \varphi_{\bf{k}} }&- \lambda_{\bf{k}}^*
 \cr  \lambda_{\bf{k}} & (2 - \lambda_{\bf{k}}) e^{- i \varphi_{\bf{k}}}\end{array} \right)        ~.
\end{equation}
Then, the final result for the momentum-space expansion of the  GW-chirality components of the Dirac spinors is:
\begin{eqnarray}
\label{psifinal}
\Psi_{- \; x} 
%&=& {1 \over 2N}\; \sum\limits_{k_1,k_2 = 1}^N \alpha_{\bf{k} -} \; \omega_N^{\bf{k}\cdot \bf{x}} \; \left[\left( \begin{array}%{c} 1 \cr e^{i \varphi_{\bf{k}}} \end{array} \right) - \left( \begin{array}{c} 1 \cr - e^{i \varphi_{\bf{k}}} \end{array} \right) \right] 
%= {1 \over N}\; \sum\limits_{\bf{k} = 1}^N \alpha_{\bf{k} -} \; \omega_N^{\bf{k}\cdot \bf{x}} \; \left( \begin{array}{c} 0 \cr e^{i %\varphi_{\bf{k}}} \end{array} \right)~\nonumber \\
&=& {1 \over N}\; \sum\limits_{\bf{k} = 1}^N \alpha_{\bf{k} -} \; \omega_N^{\bf{k}\cdot \bf{x}} \; e^{- i 
\varphi_{\bf{k}} \sigma_3 } \left( \begin{array}{c} 0 \cr 1 \end{array} \right)~, \nonumber \\
\Psi_{+ \; x} 
%{1 \over 2N}\; \sum\limits_{k_1,k_2 = 1}^N \alpha_{\bf{k} +} \; \omega_N^{\bf{k}\cdot \bf{x}} \; \left[\left( \begin{array}{c} 1 %\cr e^{i \varphi_{\bf{k}}} \end{array} \right) + \left( \begin{array}{c} 1 \cr - e^{i \varphi_{\bf{k}}} \end{array} \right) \right] 
&=&{1 \over N}\; \sum\limits_{\bf{k} = 1}^N \alpha_{\bf{k} +} \; \omega_N^{\bf{k}\cdot \bf{x}} \; \left( \begin{array}{c} 1 \cr 0 \end{array} \right)~, \\
\bar\Psi_{-\; x} 
%&=& {1 \over 2N}\; \sum\limits_{k_1,k_2 = 1}^N \bar\alpha_{\bf{k} -} \; \omega_N^{- \bf{k}\cdot \bf{x}} \;
 %\left[\left(\begin{array}{cc}1&e^{- i \varphi_{\bf{k}}}\end{array}\right) - (\lambda_{\bf{k}} - 1)  
 %\left( \begin{array}{cc} 1& - e^{-i \varphi_{\bf{k}}} \end{array} \right) \right] \nonumber \\
%&=& {1 \over 2N}\; \sum\limits_{\bf{k} = 1}^N \bar\alpha_{\bf{k} -} \; \omega_N^{-\bf{k}\cdot \bf{x}} \; \left( \begin{array}{cc} %2 - \lambda_{\bf{k}} & \lambda_{\bf{k}}\; e^{-i \varphi_{\bf{k}}} \end{array} \right)~\nonumber \\
&=& {1 \over  N}\; \sum\limits_{\bf{k} = 1}^N \bar\alpha_{\bf{k} -} \; \omega_N^{-\bf{k}\cdot \bf{x}} \; \left( \begin{array}{cc} 1 & 0 \end{array} \right) \; U_{\bf{k}}~,  \nonumber  \\
\bar\Psi_{+ \;x}
%&=& {1 \over 2N}\; \sum\limits_{k_1,k_2 = 1}^N \bar\alpha_{\bf{k} +} \; \omega_N^{- \bf{k}\cdot \bf{x}} \;
 %\left[\left( \begin{array}{cc} 1 & e^{- i \varphi_{\bf{k}}} \end{array} \right) + (\lambda_{\bf{k}} - 1)  
 %\left( \begin{array}{cc} 1& - e^{-i \varphi_{\bf{k}}} \end{array} \right) \right] \nonumber \\
%&=& {1 \over 2N}\; \sum\limits_{\bf{k} = 1}^N \bar\alpha_{\bf{k} +} \; \omega_N^{- \bf{k}\cdot \bf{x}} \; \left( \begin{array}{cc}  \lambda_{\bf{k}} &(2 - \lambda_{\bf{k}})\; e^{-i \varphi_{\bf{k}}} \end{array} \right)~
&=& {1 \over N}\; \sum\limits_{\bf{k} = 1}^N \bar\alpha_{\bf{k} +} \; \omega_N^{-\bf{k}\cdot \bf{x}} \; 
\left( \begin{array}{cc}  0 &1 \end{array} \right) V_{\bf{k}}  ~.\nonumber
\end{eqnarray}
We note again that the barred spinors are treated as two-component rows, as explicitly evident in the above expansions.
The sum in  (\ref{psifinal}) is over all momenta and thus includes also the real eigenvalues of the GW operator. 
Note also that $U_k = \sigma_3 V_{\bf{k}}^* e^{i \sigma_3 \varphi_{\bf{k}}} \sigma_3$ and that 
 the chiral projectors for the $\bar\Psi$ fields can be written in terms of  $U_{\bf{k}}$, $V_{\bf{k}}$:
\begin{equation}
\label{projectors}
\hat{P}_{\pm \; \bf{k}} = U_{\bf{k}}^\dagger \; P_\pm \;U_{\bf{k}} = V_{\bf{k}}^\dagger \; P_\pm \;V_{\bf{k}} \end{equation}
 In this notation, the fact that as $\bf{k} \rightarrow 0$ the usual and GW chirality coincide is evident---in this limit, the unitary matrices $U_{\bf{k}}, V_{\bf{k}}$ become the identity.

\mys{Fermion Green functions}
\label{fermionGFappx}
\vspace{-0.4in}

\subsection{Neutral fermion Green functions}
 
As explained in the main text, see eqn.~(\ref{SfermionNB}), we define the following fermion observable, appropriate in the large $y$ limit, when only correlators between left and right fields are nonzero: 
\begin{eqnarray}
\label{Sfermion}
S_{x-y}^n = \left( \begin{array}{cc}  \langle \chi_{+\;  x}  \; \phi_y \psi_{- \; y}^T \rangle & \langle \chi_{+\;  x}  \;   \phi^*_y \bar\psi_{- \; y}\rangle  \cr
 \langle \bar\chi_{+\;  x}^T    \; \phi_y \psi_{- \; y}^T \rangle & \langle \bar\chi_{+\;  x}^T  \;   \phi^*_y \bar\psi_{- \; y}\rangle
 \end{array} \right)
\end{eqnarray}
These correlators probe the  spectrum of neutral fermions---as explained in the Introduction, the charged fermions $\psi_-$ can bind with the charged scalars into neutral fermions, which then pair up with the neutral mirrors to form massive states; see the strong coupling expansions in 
\cite{Eichten:1985ft, Hasenfratz:1988vc} for discussion of analogous phenomena.

Then define, for every value of $k$, the four-by-four matrix propagator in momentum space, and use the expressions (\ref{psifinal}) ($\chi_+$ is obtained by replacing $\alpha_+ \rightarrow \beta_+$ in the corresponding equation in (\ref{psifinal})): 
\begin{eqnarray}
\label{Sfermion2}
s_{\bf k}^n &=& \sum_x \omega_N^{- {\bf k} (x - y)} S^n_{x - y} \\
&=&\sum\limits_{\bf p} \left( \begin{array}{cc} \langle \beta_{\bf k} \alpha_{\bf p} \Phi_{{\bf -k-p}} \rangle \;  e^{i \varphi_{\bf p}} \sigma_+ & \langle \beta_{\bf k} \bar\alpha_{{\bf k+p}} \Phi^*_{\bf p}  \rangle \;  P_+ U_{{\bf k+p}}  \cr \langle \bar\beta_{{\bf -k}}  \alpha_{{\bf p-k}} \Phi_{{\bf -p}} \rangle \; V_{{\bf -k}}^T P_- e^{ i \varphi_{{\bf p-k} }} &  \langle \bar\beta_{{\bf -k}} \bar\alpha_{\bf p} \Phi^*_{{\bf p-k}} \rangle \; V_{{\bf -k}}^T \sigma_- U_{\bf p }
\end{array} \right)~.\nonumber
\end{eqnarray}
Here, (\ref{unitaryprojectors}) defines the $U$ and $V$ matrices and $e^{i \varphi}$ is  defined  in (\ref{phi}). We will study numerically the following quantity:
\begin{eqnarray}
\label{omega1}
\Omega^{(1)}_{\bf{k}} = \frac{1}{ \sqrt{ {\rm Tr}\; s_{\bf{k}}^{n \;\dagger} s_{\bf{k}}^n}}~,
\end{eqnarray}
which, as discussed in Section \ref{chiralmassless},  provides a lower bound on the smallest neutral fermion inverse propagator eigenvalue. 

 It is instructive to consider the form of the neutral Green functions for constant $\phi_x = 1$ background, where they can be evaluated explicitly. This is, of course, only relevant for large values of $\kappa$ where the fluctuations of $\phi_x$ are nearly frozen. Note that for $\phi_x = 1$ there is no difference between the charged and neutral Green functions (defined in following section).
 The propagator $s_{\bf{k}}$ then takes the form:
\begin{eqnarray}
\label{constantphiS}
s_{\bf k}^0 = \left( \begin{array}{cc} \langle \beta_{\bf k} \alpha_{\bf - k} \rangle \sigma_+ e^{i \varphi_{- {\bf k}}} & \langle \beta_{\bf k} \bar\alpha_{\bf k} \rangle P_+ \; U_{\bf k} \cr \langle \bar\beta_{- \bf{k}} \alpha_{- \bf{k}} \rangle V_{- \bf{k}}^T P_- e^{i \varphi_{- {\bf k}}} & \langle \bar\beta_{- \bf{k}} \bar\alpha_{\bf{k}} \rangle V_{- {\bf k}}^T \sigma_-  U_{ \bf k} \end{array} \right) ~.
\end{eqnarray}
The biunitary transform of $s_{\bf k}^0$:
\begin{eqnarray}
\label{constantphiS1}
\hat{s}_{\bf k}^0 \equiv \left(\begin{array}{cc} 1&0 \cr 0 &   V_{- \bf{k}}^* \end{array}  \right)  \; s_{\bf{k}}^0  \left(\begin{array}{cc} 1 &0 \cr 0 & U_{\bf{k}}^\dagger \end{array}  \right)~,
\end{eqnarray} takes the even  simpler form:
\begin{eqnarray}
\label{constantphiS2}
\hat{s}_{\bf k}^0 = \left( \begin{array}{cc} \langle \beta_{\bf k} \alpha_{\bf - k} \rangle \sigma_+ e^{i \varphi_{- {\bf k}}} & \langle \beta_{\bf k} \bar\alpha_{\bf k} \rangle P_+   \cr \langle \bar\beta_{- \bf{k}} \alpha_{- \bf{k}} \rangle P_- e^{i \varphi_{- {\bf k}}} & \langle \bar\beta_{- \bf{k}} \bar\alpha_{\bf{k}} \rangle   \sigma_-    \end{array} \right) 
\end{eqnarray}
and makes explicit the fact that the rank of the matrix is two, the number of propagating complex degrees of freedom.
The relevant expectation values entering the Green functions can be easily computed for constant $\phi_x$ and are given by (it is important to take into account the measure factor $1-\lambda_{\bf k}^*$ (\ref{mirrormeasure}) when calculating the expectation values): 
\begin{eqnarray}
\label{constantphi}
\langle \beta_{\bf k} \alpha_{\bf - k} \rangle &=& - i h e^{ - i \varphi_{\bf k}} {4 - 4 \lambda_{\bf k} + 2 \lambda_{\bf k}^2 \over 4} \; d({\bf k})^{-1} \\
\langle \beta_{\bf k} \bar\alpha_{\bf k} \rangle &=& { 2 - \lambda_{\bf k}^* \over 2 } \; d({\bf k})^{-1} \nonumber \\
\langle \bar\beta_{\bf -k} \bar\alpha_{\bf k} \rangle &=& - i h e^{ i \varphi_{\bf k}} (1 - \lambda_{\bf k}^*) \;  d({\bf k})^{-1} \nonumber \\
\langle \bar\beta_{-\bf k} \alpha_{\bf - k} \rangle &=& - {2 - \lambda_{\bf k} \over 2} \; d({\bf k})^{-1}~, \nonumber 
\end{eqnarray}
where 
\begin{eqnarray}
\label{Dk}
d({\bf k}) = \sin^2 {\theta_{\bf k} \over 2} + h^2 \cos \theta_{\bf k}, ~~ \lambda_{\bf k} \equiv 1 + e^{i \theta_{\bf k}} ~.
\end{eqnarray}
Combining (\ref{constantphiS2}, \ref{constantphi}) we find:
\begin{eqnarray}
\label{constantphiS3}
\hat{s}_{\bf k}^0 = {1\over d(\bf{k})}\left( \begin{array}{cc} - i h\; {4 - 4 \lambda_{\bf k} + 2 \lambda_{\bf k}^2 \over 4}  \; e^{i (\varphi_{- {\bf k}}- \varphi_{\bf k})} \; \sigma_+ &  { 2 - \lambda_{\bf k}^* \over 2 } \; P_+   \cr - {2 - \lambda_{\bf k} \over 2}\;  e^{i \varphi_{- {\bf k}}} \; P_-  & - i h \;e^{ i \varphi_{\bf k}} (1 - \lambda_{\bf k}^*) \; \sigma_-    \end{array} \right) ~.
\end{eqnarray}
The rank of $\hat{s}_{\bf k}^0$ is two; clearly, 
the nontrivial part is obtained by simply replacing 
$\sigma_\pm$ and $P_\pm$ by  unity.  After the elimination
of two rows and columns of zeros, the determinant 
of $\hat{s}_{\bf k}^0$ equals $e^{i \varphi_{\bf -k}} d({\bf k})^{-1}$.  
Clearly, the denominator  (\ref{Dk}) of $\hat{s_k^0}$ admits 
small eigenvalues for values of $\bf{k}$ where (assuming $h>1/2$):
\begin{equation}
\label{smalleigenvalues}
\big\vert \sin {\theta_{\bf{k}}\over 2} \big\vert = {h \over \sqrt{2 h^2 - 1}}~.
\end{equation}
  For generic values of $h$ this equation has no exact solutions at finite $N$, except for $h=0$, where 
  the $\theta_{\bf k} = 0$ corresponds to the three unlifted modes with $\lambda_{\bf k} = 2$, and $h=1$, corresponding to $\theta_{\bf k}  = \pi$, where the ${\bf k} = (N,N)$ mode is unlifted.  Other values of $\bf{k}$ for which (\ref{smalleigenvalues})  approximately holds occur,   for sufficiently large $h$ near $\theta_{\bf{k}} \simeq {\pi \over 2}$, and give rise to small mass eigenvalues in the broken phase.
 
As explained in the main text below \myref{diracmirror},
the $h=1$ exact zero mode persists for small $\kappa$ 
as well.  It gives rise to the critical behavior seen in our
numerical analysis at $h \approx 1$, and discussed 
in the main body of the paper.

As a consistency check on our numerical simulations, we have also verified that for large values of $\kappa$, in the algebraically ordered phase, where   perturbation theory leading to (\ref{constantphiS3}) is valid, the mass matrix, including the small eigenvalues due to  (\ref{smalleigenvalues}), are reproduced by our Monte-Carlo simulations. 

Most importantly, however, as described in the main text, in the disordered phase with rapid fluctuations of $\phi_x$, our results for $\Omega^{1,2}_{\bf{k}}$ show that there are no small eigenvalues
in the $h > 1$ regime.

\subsection{Charged fermion Green functions}

The correlators that probe the charged fermion spectrum, eqn.~(\ref{SfermionChB}), where charged bound states of the neutral mirrors $\chi_+$ and the scalars can pair up with the charged mirrors to form heavy fermions, appropriate to the large-$y$ limit are as follows: 
\begin{eqnarray}
\label{SfermionCh}
S^{c }_{y-x} = \left( \begin{array}{cc}  \langle \psi_{- \; x}  \; \chi_{+\;  y}^T    \phi_y  \rangle &
 \langle \chi_{+\;  x}   \phi_x ^*\; \bar\psi_{- \; y} \rangle \cr
  \langle  \psi_{- \; x} \; \bar\chi_{+\;  y}   \phi_y \rangle  & 
\langle \bar\chi_{+\;  x}^T     \phi_x^* \; \bar\psi_{- \; y}  \rangle 
 \end{array} \right)~ \;
\end{eqnarray}
For every value of $k$, the four-by-four matrix propagator in momentum space is, then, similar to (\ref{Sfermion2}): 
\begin{eqnarray}
\label{SfermionCh2}
s_{\bf k}^{c} &=& \sum_x \omega_N^{- {\bf k} (x - y)} S^c_{x - y} \\
&=&\sum\limits_{\bf p} \left( \begin{array}{cc} \langle \alpha_{\bf k} \beta_{\bf p}  \Phi_{{\bf -k-p}} \rangle \;  e^{i \varphi_{\bf k}} \sigma_- &
 \langle \beta_{\bf  p} \bar\alpha_{{\bf k }} \Phi^*_{\bf k-p}  \rangle \;  P_+ U_{{\bf k}}  \cr 
 \langle \alpha_{{\bf k}} \bar\beta_{{\bf p}}   \Phi_{{\bf p-k}} \rangle \;  P_- V_{{\bf p}}  e^{ i \varphi_{{\bf k} }} &  \langle \bar\alpha_{\bf k} \bar\beta_{{\bf p}}  \Phi^*_{{\bf p+k}} \rangle \; V_{{\bf p}}^T \sigma_- U_{\bf k }
\end{array} \right)~ .\nonumber
\end{eqnarray}
The quantity that we numerically study is defined as for the neutral propagator (\ref{omega1}):
\begin{eqnarray}
\label{omega2}
\Omega^{(2)}_{\bf{k}} = \frac{1}{\sqrt{ {\rm Tr}\; s_{\bf{k}}^{c \;\dagger} s_{\bf{k}}^c}}~,
\end{eqnarray}
and similarly provides a lower bound on the smallest eigenvalue of the inverse charged-fermion  propagator.

\subsection{Fermion-bilinear susceptibilities}
\label{fermionSUSC}

There are two composite charged complex scalars that we can construct out of bilinears of the mirror fermions---$  \bar\psi_{- \; x} \chi_{+ \; x} $ and $\psi_{- \; x}^T \gamma_2 \chi_{+ \; x}$.  Consider the corresponding susceptibilities:
\begin{eqnarray}
\label{chargedcompositescalar}
\Delta_{x - y} &\equiv& \langle \bar\psi_{- \; x} \chi_{+ \; x} \; \bar\chi_{+ \; y} \psi_{- \; y} \rangle \\
 &=& {1\over N^4 } \sum\limits_{\bf k,p,q,l} \langle \bar\alpha_{\bf k} \; \beta_{\bf p} \; \bar\beta_{\bf q}  \; \alpha_{\bf l} \rangle \; \omega_N^{({\bf p - k}) x} \; \omega_N^{({\bf l - q}) y}  ~ { (2 - \lambda_{\bf{k}}) (2 - \lambda_{\bf{q}}) \over 4} ~ e^{ i( \varphi_{\bf{l} } - \varphi_{\bf{q}})}~. 
\end{eqnarray}
We now define a susceptibility as in the scalar case:
\begin{eqnarray}
\label{fermionsuscept}
\chi_F &\equiv& \sum_x \Delta_{x - y}\vert_{connected}\\
&=& {1\over N^2} \sum\limits_{\bf k,q}  { (2 - \lambda_{\bf{k}}) (2 - \lambda_{\bf{q}}) \over 4} 
\left( \langle \bar\alpha_{\bf k} \; \beta_{\bf k} \; \bar\beta_{\bf q}  \; \alpha_{\bf q} \rangle  -  \langle \bar\alpha_{\bf k} \; \beta_{\bf k} \rangle \langle \bar\beta_{\bf q}  \; \alpha_{\bf q} \rangle\right) ~,
\end{eqnarray}
where we subtracted the disconnected component, which should vanish anyway in the symmetric phase (we have checked that it, indeed, does vanish).
Consider also the "Majorana" correlator:
\begin{eqnarray}
\label{chargedcompositescalar2}
\Delta^\prime_{x-y} &\equiv&  \langle \psi_{- \; x}^T \gamma_2 \chi_{+ \; x} \; \bar\chi_{+ \; y} \gamma_2 \bar\psi_{- \; y}^T \rangle \\
&=&-{1 \over 4  N^4}\sum\limits_{\bf k,p,q,l} \langle \alpha_{\bf k} \beta_{\bf p} \; \bar\beta_{\bf q} \bar\alpha_{\bf l} \;  \rangle e^{i \varphi_{\bf k}} \; \omega_N^{({\bf k + p}) x}\;  \omega_N^{- ({\bf q + l}) y} \left( (2- \lambda_{\bf q}) (2 - \lambda_{\bf l}) e^{ - i \varphi_{\bf q}}  - \lambda_{\bf q} \lambda_{\bf l} e^{- i \varphi_{\bf l}} \right)~, \nonumber
\end{eqnarray}
and the corresponding "Majorana susceptibility:"
\begin{eqnarray}
\label{fermionsuscept2}
\chi_F^\prime &\equiv& \sum_x \Delta_{x - y}^\prime\vert_{connected} \\
&=& -{1\over 4 N^2} \sum\limits_{\bf k,q}   e^{i (\varphi_{\bf k} -   \varphi_{\bf q})} \left( 4 - 4 \lambda_{\bf q} + 2 \lambda_{\bf q}^2 \right) \left(  \langle \alpha_{\bf k} \beta_{\bf -k} \; \bar\beta_{\bf q} \bar\alpha_{\bf -q} \rangle - \langle \alpha_{\bf k} \beta_{\bf -k}\rangle \langle \bar\beta_{\bf q} \bar\alpha_{\bf -q} \rangle\right)~, \nonumber 
\end{eqnarray}
where, again, we subtracted the disconnected part, which (as we checked, once more) vanishes in the symmetric phase.

\mys{Simulation details}
\label{simdetails}
The configurations of $XY$ fields $\eta$ are generated using the
Wolff single-cluster algorithm \cite{Wolff:1988uh}, which is well-known to
overcome critical slowing down for this lattice system.
The fermion measure is taken into account through
determinant reweighting:
\beq
\vev{\Ocal} = \frac{ \vev{\Ocal \det M}_\eta }{\vev{ \det M}_\eta}.
\eeq
Here, $\Ocal$ is any observable, and $\vev{ \cdots }_\eta$
denotes an expectation value with respect to 
the measure of the XY model (cf.~eq.~\myref{Skappa} with $U \equiv 1$), 
\beq
d\mu(\eta) = Z_{XY}^{-1} \(\prod_x d\eta_x\) \exp (-S_\kappa).
\eeq
We monitor the reliability
of this method in three ways.
\begin{enumerate}
\item We measure the
autocorrelation time for reweighted quantities
$\vev{\Ocal \det M}_\eta$, as
well as $\vev{\det M}_\eta$, to be certain that the configurations
remain independent with respect to the new measure.
\item We perform a jackknife error analysis of
the averages $\vev{\Ocal}$ that are obtained, gathering sufficient
data to keep errors small.  (An ``overlap problem'' would
be indicated by large errors.)  \item We have simulated
a number of sample points in parameter space by
an alternative method that does not rely on
determinant reweighting.  Namely, we have used Metropolis
updates that include $\Delta S = -\ln \det M$ in
the action.  We check that this alternative
(significantly slower, due to much longer
autocorrelation times that occur when using the Metropolis
algorithm) method leads to results
that agree with the determinant reweighting method.
Sample comparison points are presented in Table~\ref{cmpt}.
\end{enumerate}
We find that the Wolff cluster algorithm together
with determinant reweighting is efficient and
reliable for all quantities and regions of
parameter space that we have explored.  The only
difficulty that occurs is a sign problem in the
$h<1$ regime, as mentioned in the main text.
However, this is easy to detect due to the
associated large statistical errors.

\begin{table}
\begin{center}
\begin{tabular}{ccccc}
$N$ & $h$ & $\kappa$ & Metropolis & Wolff + det RW \\ \hline
4 & 4 & 0.1 & 1.423(5) & 1.424(8) \\
4 & 2 & 0.5 & 3.65(4) & 3.57(4) \\
8 & 4 & 0.1 & 1.026(4) & 1.021(8) \\ \hline
\end{tabular}
\caption{Sample comparison points, Metropolis results versus Wolff algorithm
with determinant reweighting. \label{cmpt} }
\end{center}
\end{table}

\subsection{Comments on $y^{-1} \not=0$}
\label{finitey}
To include $y< \infty$ in our simulations, 
we have to work with a Pfaffian, as is clear from 
the results of Section (\ref{splitofZ}), 
eqns.~(\ref{mirrorkinetic}, \ref{diracmirror}).
A method to compute the Pfaffian, including the phase,
is found in the appendix of ref.~\cite{Catterall:2003ae}. 
We want to determine the sign ambiguity
that occurs in $\Pf {\cal M}= \pm \det {\cal M}^{1/2}$,
since it is obviously crucial to any averaging; 
here, we denoted by ${\cal{M}}$ the fermion 
matrix determined by eqns.~(\ref{mirrorkinetic}, \ref{diracmirror}) 
at $y<\infty$. Similarly, we denote  by $M$ the fermion 
matrix in the $y\rightarrow \infty$ limit.

For $y^{-1} \lappeq 10^{-3}$, we find that $\ln \Pf {\cal M} = \ln \det M$
to within 4 or 5 digits for $N=4$ and $N=8$ lattices.
That is so close that it will
not change any of the results that were 
stated in Sections (\ref{strongsymmetric}, \ref{chiralmassless}) above.
The result $\ln \Pf {\cal M} = \ln \det M$
does not appear to depend on the lattice size. 

Right at $h=1$,  $M$ becomes singular for arbitrary $\phi$ backgrounds and 
there is an increased sensitivity to the $y^{-1}$ corrections.

Returning to $h \not=1$, here are some more empirical results.
For $y^{-1} \gappeq 0.1$, $\Pf {\cal M}$ is
noticably different from $\det M$.  However, the complex
phase seems much less sensitive to $y^{-1}$ 
corrections than the magnitude.  The complex phase
of $\Pf {\cal M}$ does fluctuate a bit from
configuration to configuration.  But for $y=10$,
it is a fluctuation in the fourth significant
digit---hardly a ``complex phase problem.''

\vspace{10cm}
  
\begin{figure}
\begin{center}
\includegraphics[width=3in,height=5in,angle=90]{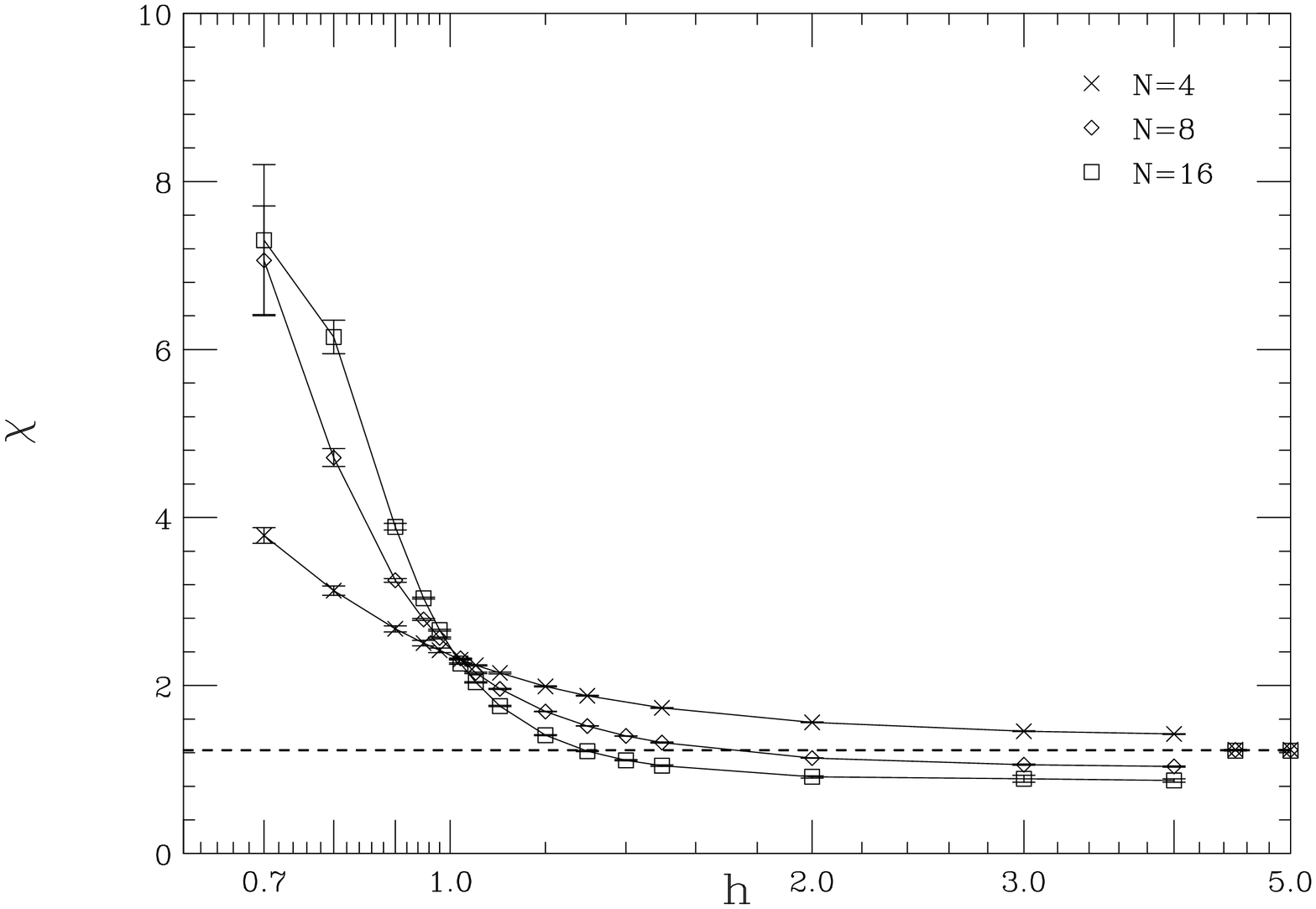}
\caption{\small Comparison of susceptibilities for $\kappa=0.1$. The dashed line indicates the susceptibilities for the pure $XY$ model with same $\kappa$ (undistinguishable, within errors for $N=4,8,16$).
Here and below, large errors at $h =$ 0.7 and 0.8
are due to the sign problem at $h<1$.}\label{fig:kappa0.1scalar}
\end{center}
\end{figure}

\begin{figure}
\begin{center}
\includegraphics[width=3in,height=5in,angle=90]{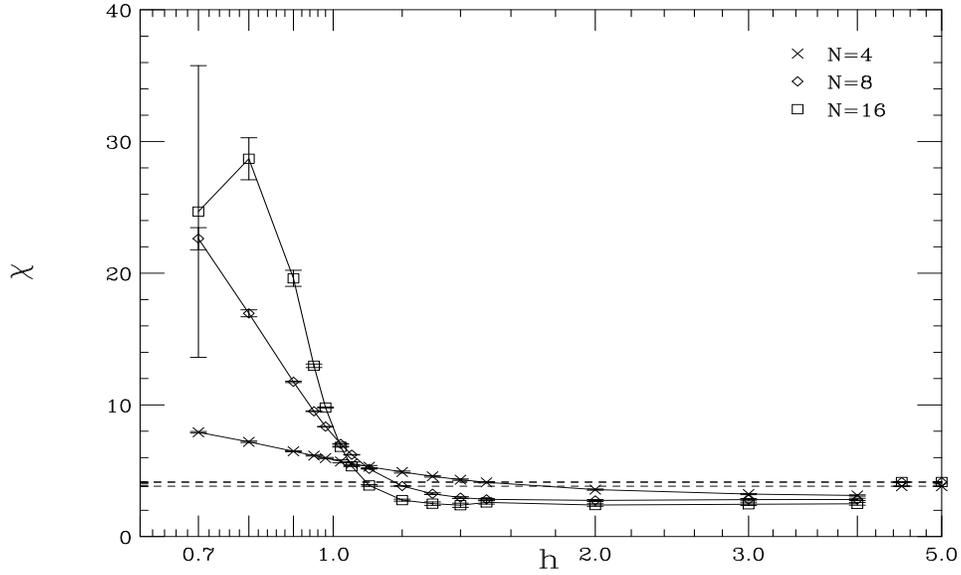}
\caption{\small Comparison of susceptibilities for $\kappa=0.5$. The dashed lines indicate  the susceptibilities for the pure $XY$ model with same $\kappa$.}\label{fig:kappa0.5scalar}
\end{center}
\end{figure}

\begin{figure}
\begin{center}
\includegraphics[width=3in,height=5in,angle=90]{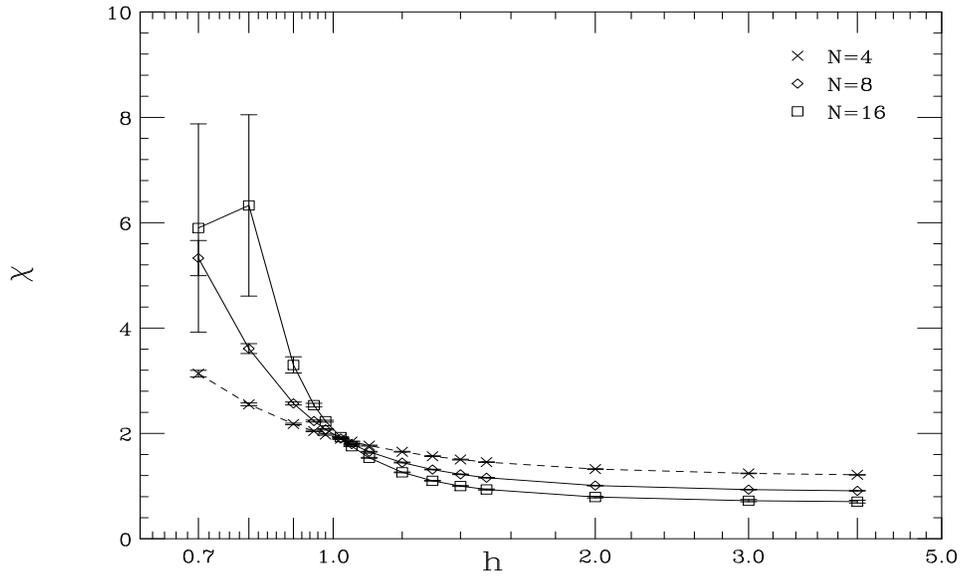}
\caption{\small Comparison of susceptibilities for $\kappa=0$.
The results are qualitatively similar to those obtained 
for $\kappa = 0.1, 0.5$. }\label{fig:kappazero}
\end{center}
\end{figure}

\begin{figure}
\begin{center}
\includegraphics[width=3in,height=5in,angle=90]{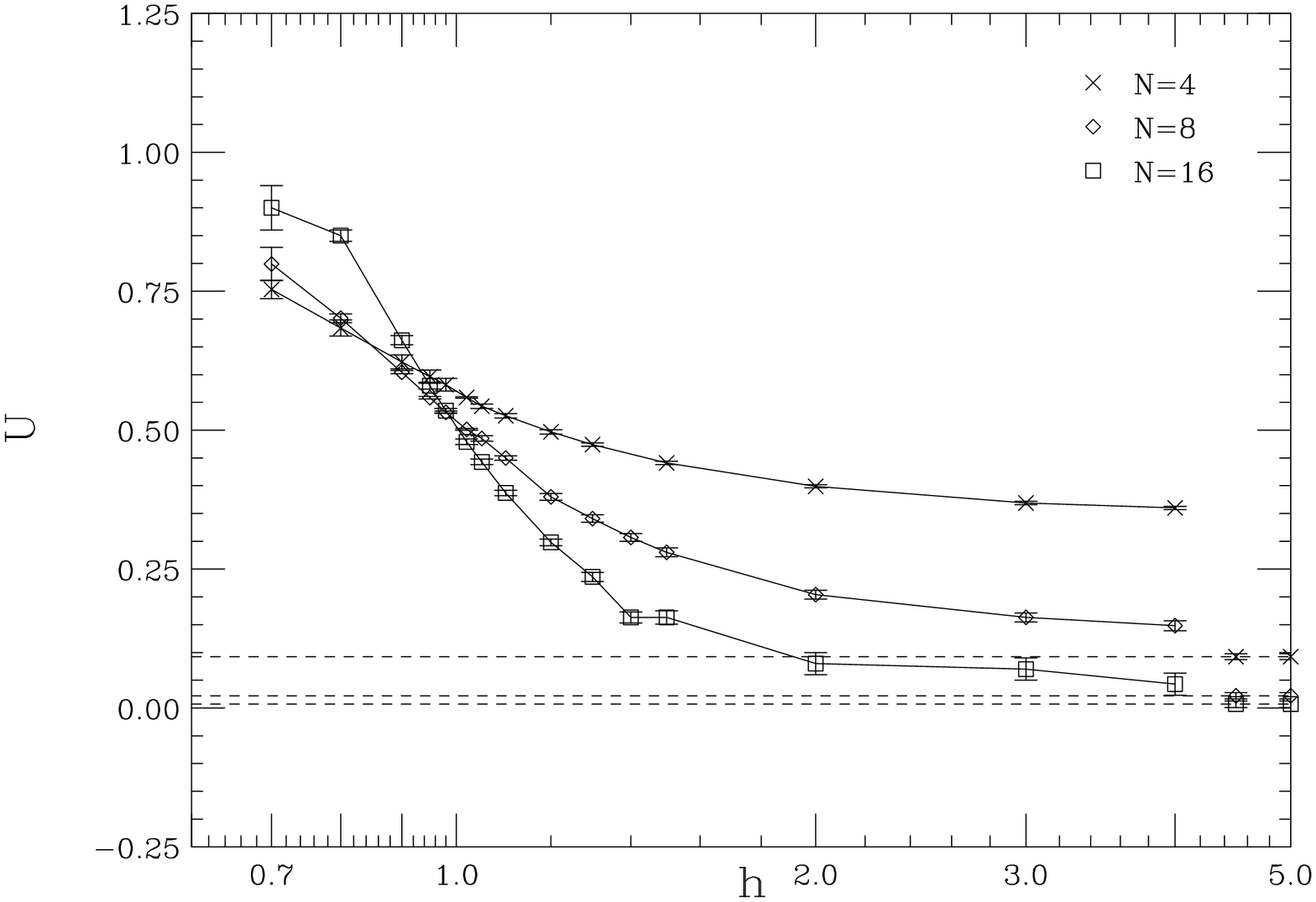}
\caption{\small Comparison of Binder cumulant for $\kappa=0.1$. The dashed lines indicate  the values of the Binder cumulant for the pure $XY$ model with same $\kappa$. }\label{fig:kappa0.1binder}
\end{center}
\end{figure}

\begin{figure}
\begin{center}
\includegraphics[width=3in,height=5in,angle=90]{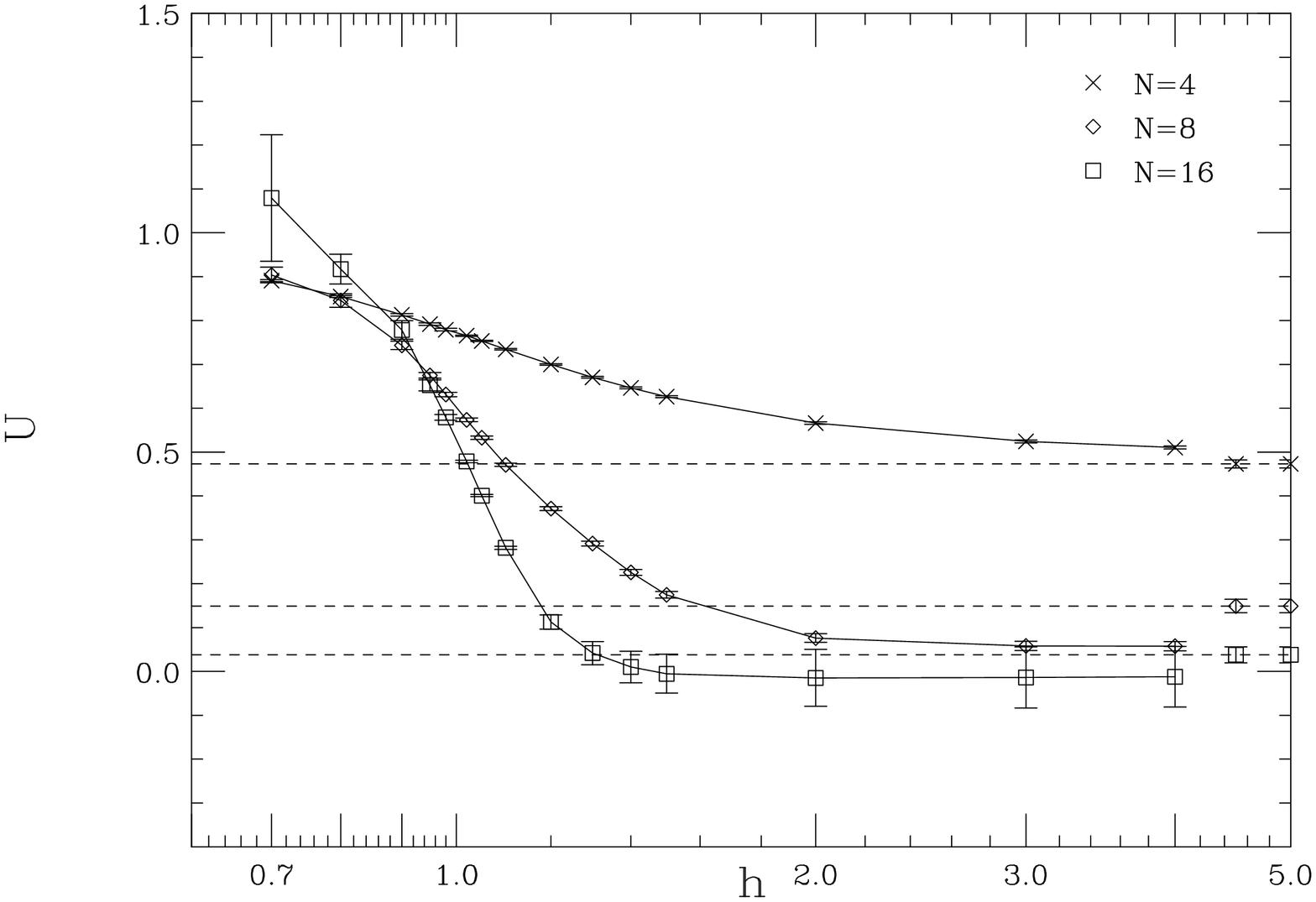}
\caption{\small Comparison of Binder cumulant for $\kappa=0.5$. The dashed lines indicate  the values of the Binder cumulant for the pure $XY$ model with the same $\kappa$.}\label{fig:kappa0.5binder}
\end{center}
\end{figure}

\begin{figure}
\begin{center}
\includegraphics[width=3in,height=5in,angle=90]{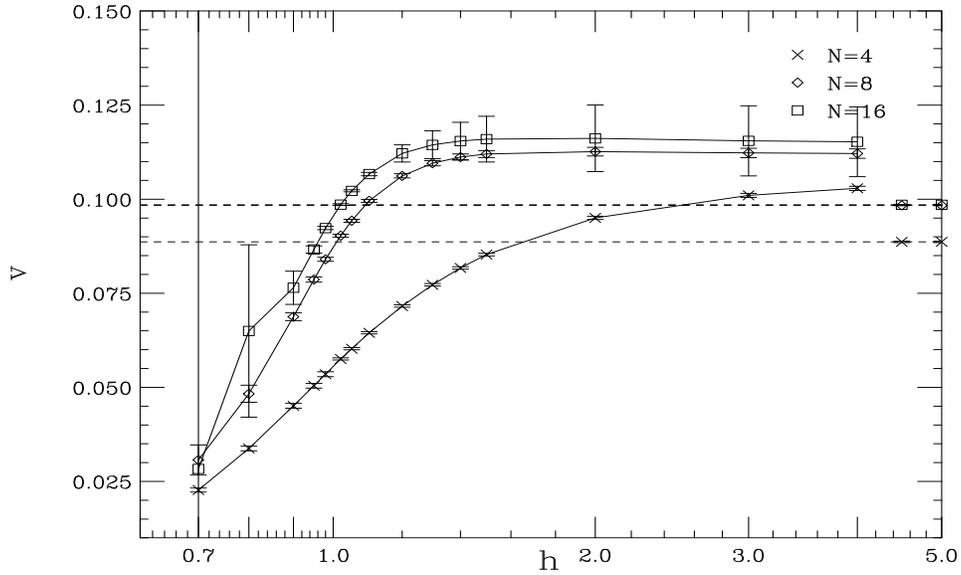}
\caption{\small The vortex density for $\kappa=0.5$ as a function of $h$.   The pure $XY$ model values of the vortex densities for $\kappa = 0.5$ are shown in the corresponding dashed lines. The slight increase of the vortex density at $h>1$ (especially for $N = 8,16$) 
implies that   the fermions  lower the vortex energy.  }\label{fig:kappa0.5vortex}
\end{center}
\end{figure}

\begin{figure}
\begin{center}
\includegraphics[width=3in,height=5in,angle=90]{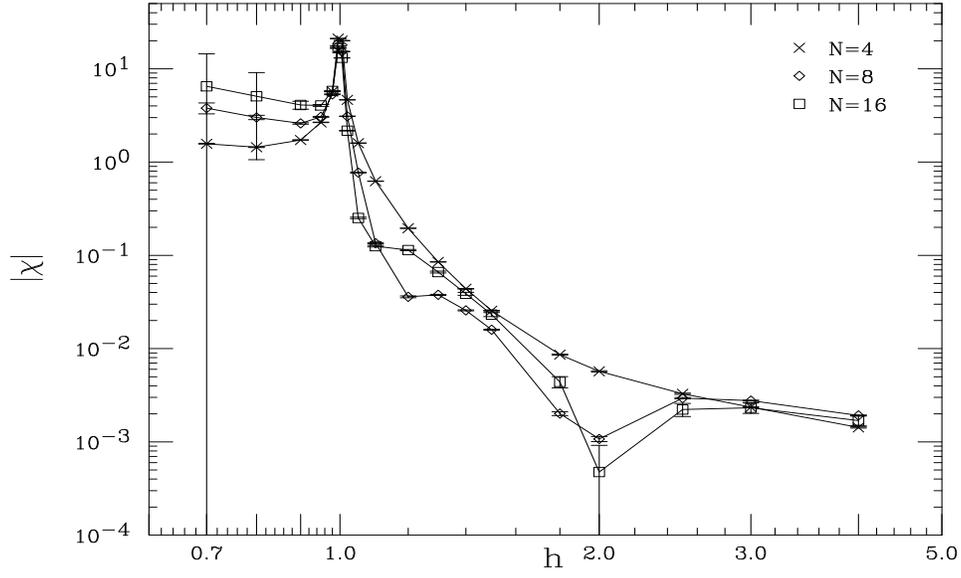}
\caption{ \small The magnitude of the ``Dirac'' susceptibility $\chi_F$ for $\kappa = 0.5$. The values of all physical 
susceptibilities are obtained from the plot by multiplying the 
plotted value by the inverse square power of the dimensionless Yukawa  coupling $1/Y^2 = 1/(ay)^2$.}
\label{fig:chiF1} 
\end{center}
\end{figure}

\begin{figure}
\begin{center}
\includegraphics[width=3in,height=5in,angle=90]{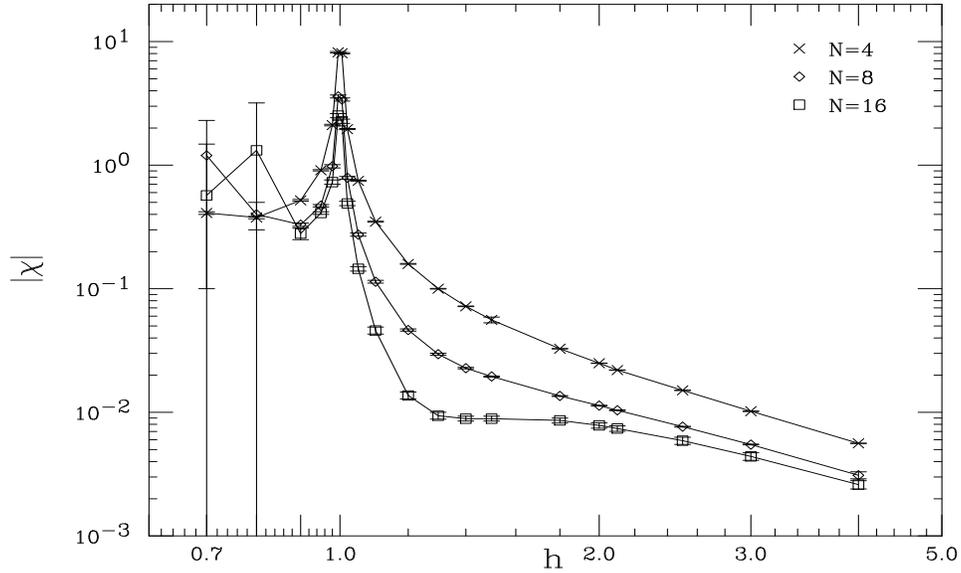}
\caption{ \small The magnitude of the ``Dirac'' susceptibility $\chi_F$ for $\kappa = 0.1$.}
\label{fig:chiF10.1} 
\end{center}
\end{figure}

\begin{figure}
\begin{center}
\includegraphics[width=3in,height=5in,angle=90]{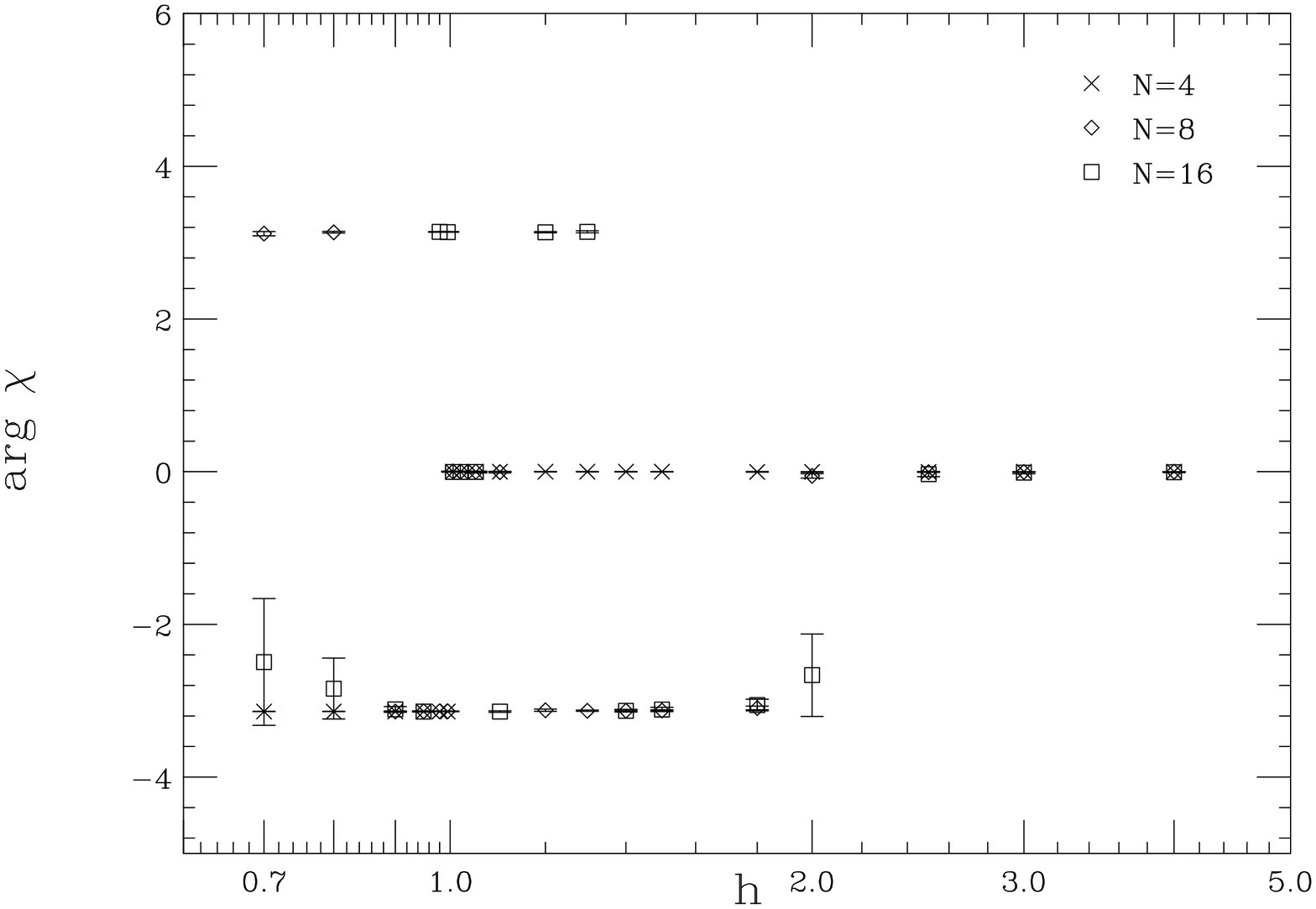}
\caption{ \small The argument of the ``Dirac'' susceptibility $\chi_F$ for $\kappa = 0.5$.}
\label{fig:chiF2} 
\end{center}
\end{figure}

\begin{figure}
\begin{center}
\includegraphics[width=3in,height=5in,angle=90]{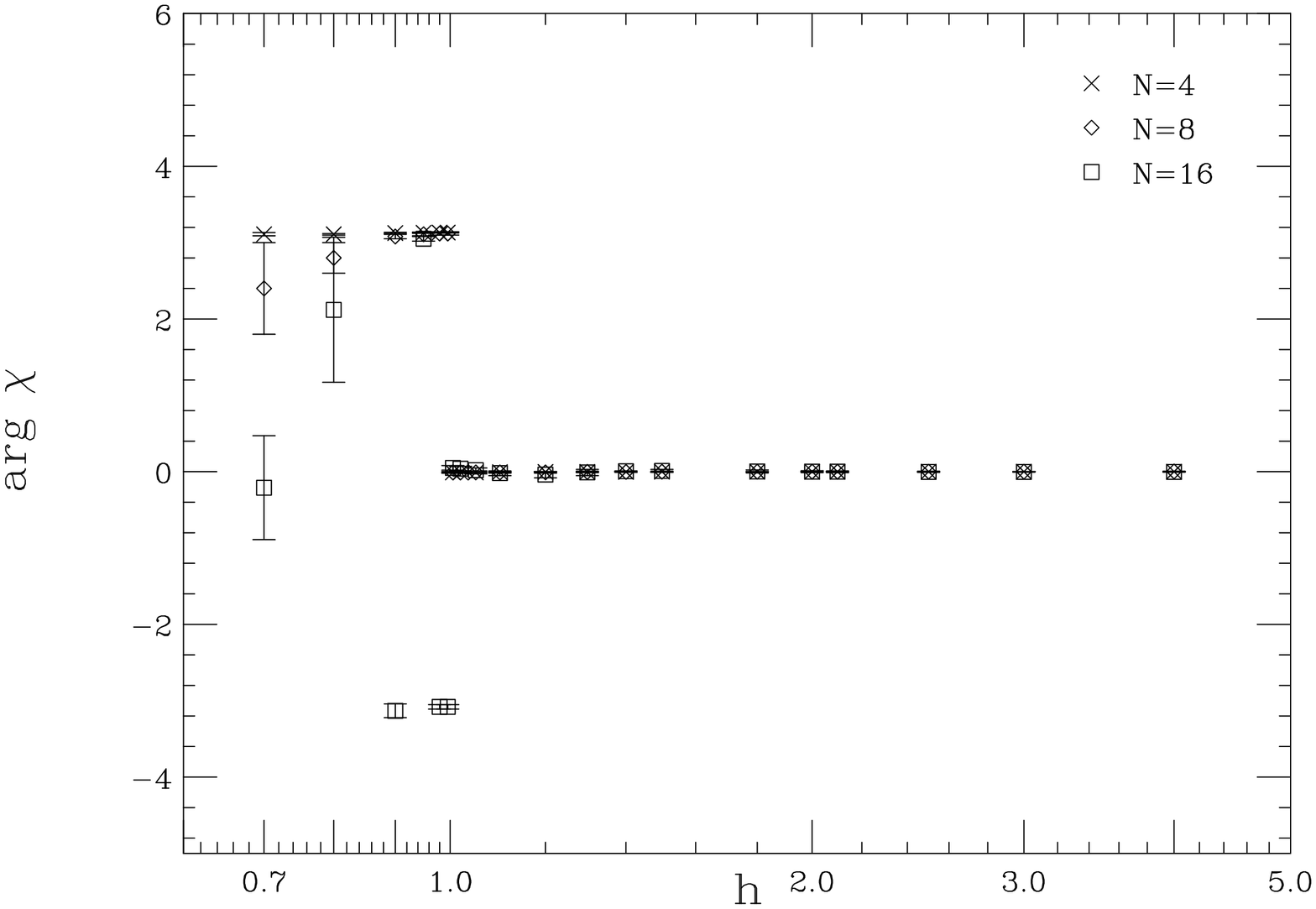}
\caption{ \small The argument of the ``Dirac'' susceptibility $\chi_F$ for $\kappa = 0.1$.}
\label{fig:chiF20.1} 
\end{center}
\end{figure}

\begin{figure}
\begin{center}
\includegraphics[width=3in,height=5in,angle=90]{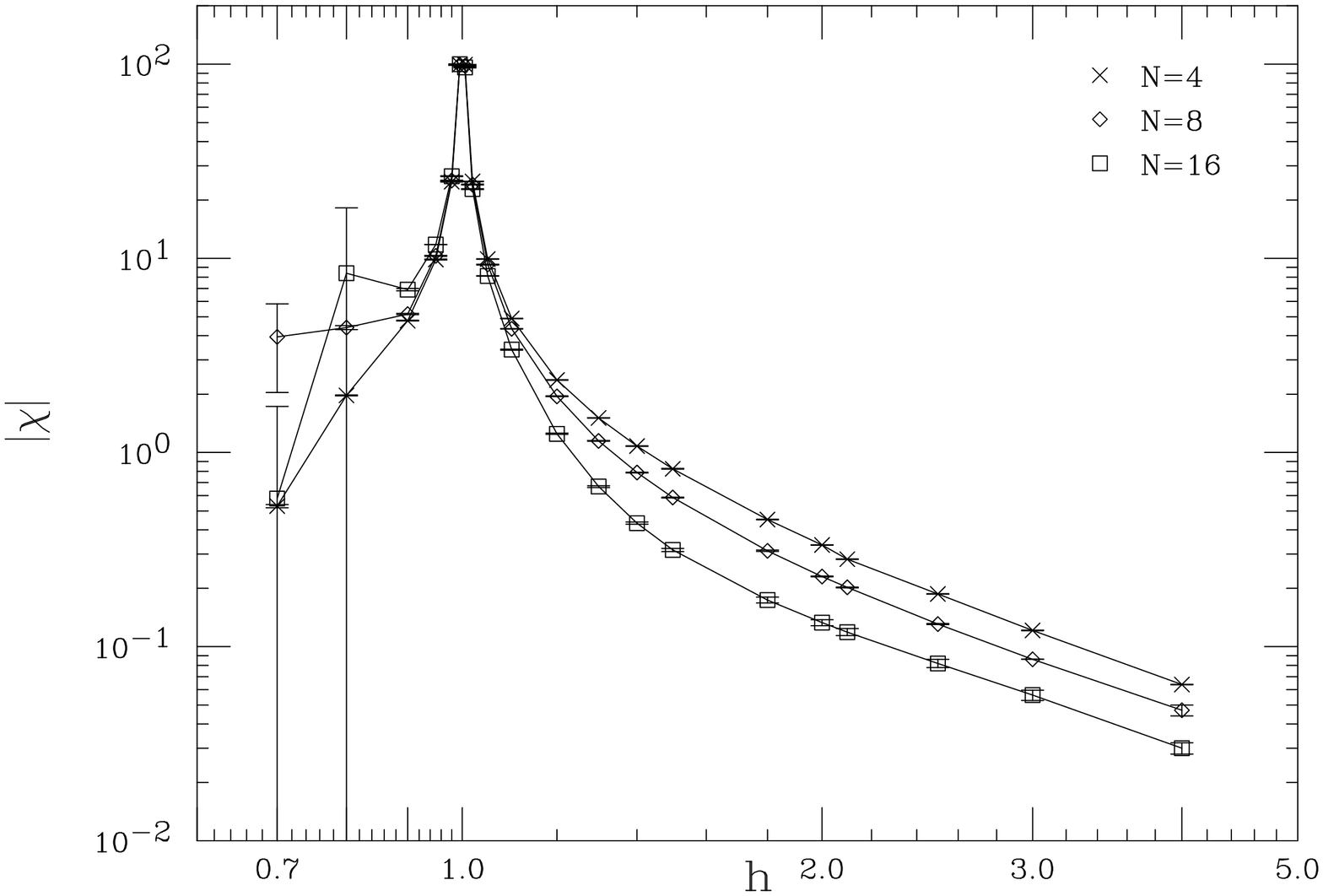}
\caption{\small The magnitude of the ``Majorana'' susceptibility $\chi'_F$ for $\kappa = 0.1$.}
\label{fig:chiFp10.1} 
\end{center}
\end{figure}

\begin{figure}
\begin{center}
\includegraphics[width=3in,height=5in,angle=90]{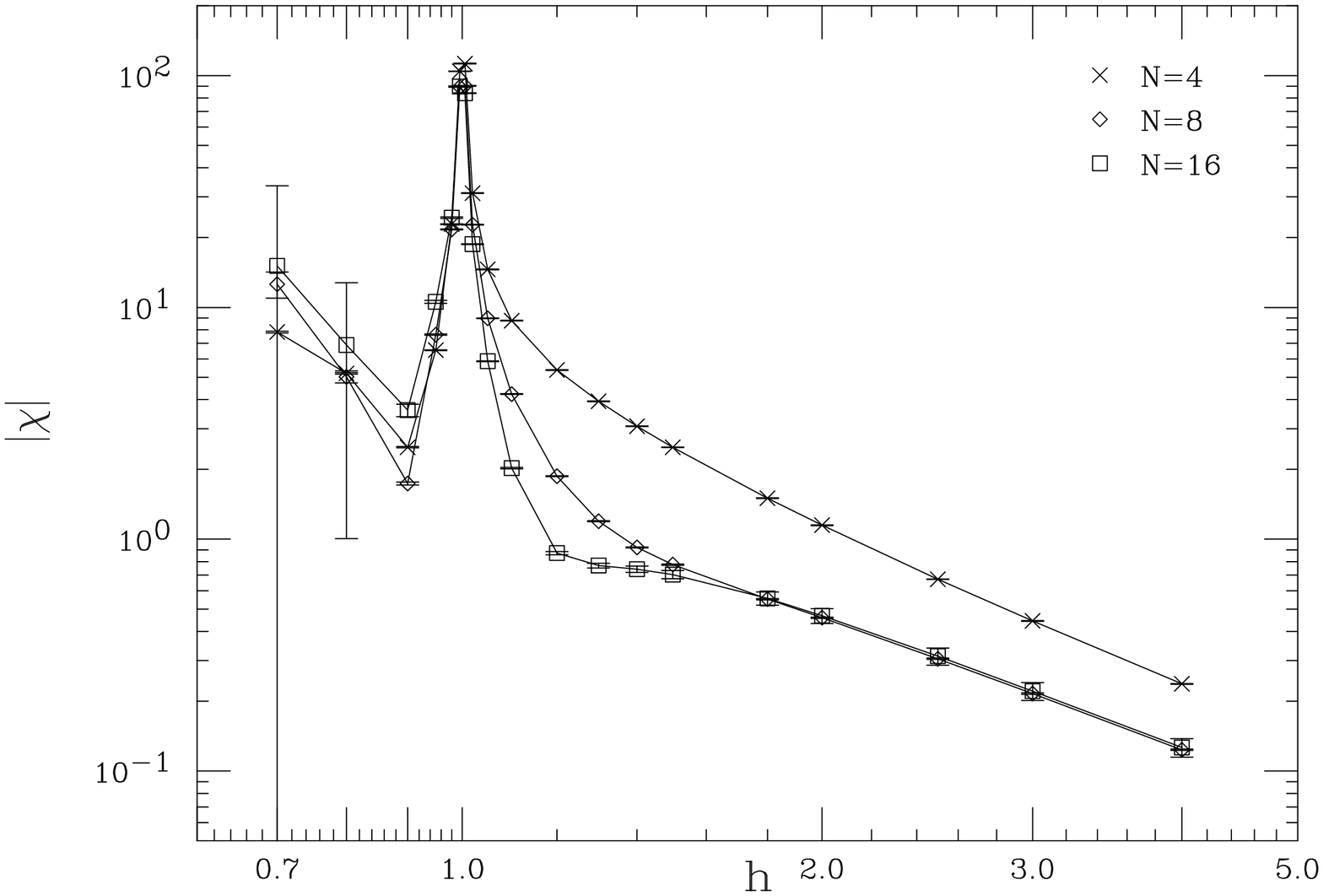}
\caption{\small The magnitude of the ``Majorana'' susceptibility $\chi'_F$ for $\kappa = 0.5$.}
\label{fig:chiFp1} 
\end{center}
\end{figure}

\begin{figure}
\begin{center}
\includegraphics[width=3in,height=5in,angle=90]{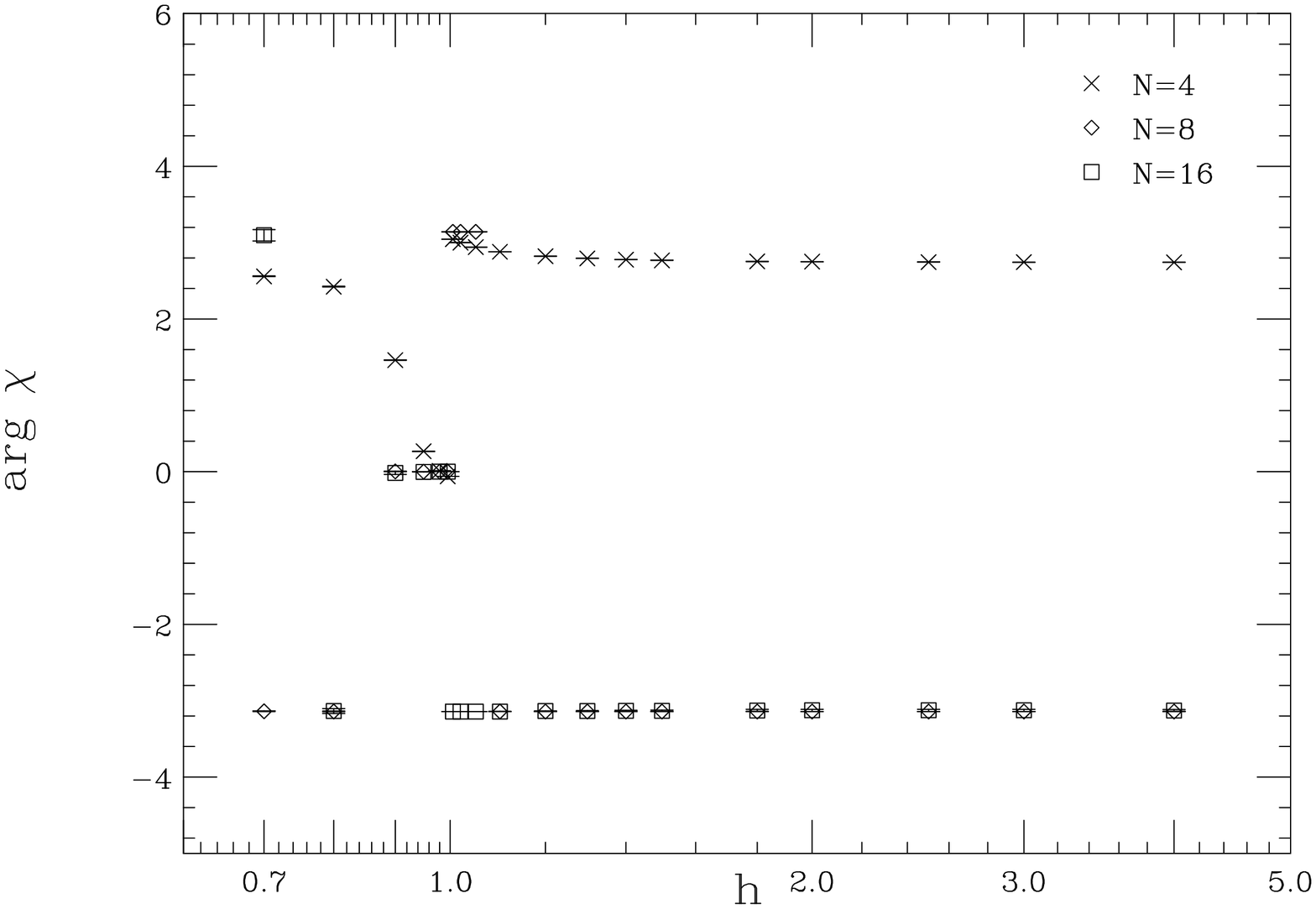}
\caption{\small The argument of the ``Majorana'' susceptibility $\chi'_F$ for $\kappa = 0.5$.}
\label{fig:chiFp2} 
\end{center}
\end{figure}

\begin{figure}
\begin{center}
\includegraphics[width=3in,height=5in,angle=90]{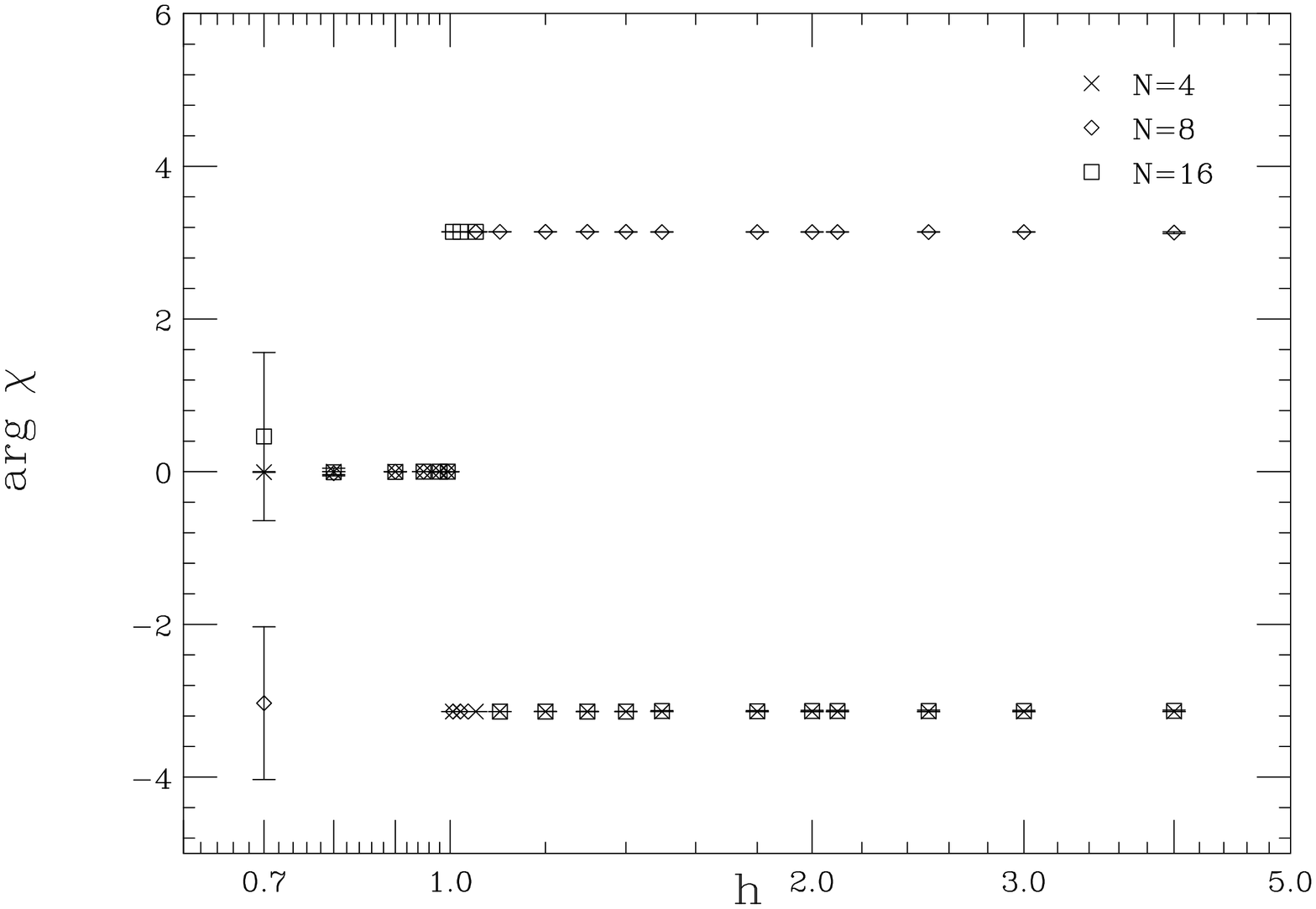}
\caption{\small The argument of the ``Majorana'' susceptibility $\chi'_F$ for $\kappa = 0.1$.}
\label{fig:chiFp20.1} 
\end{center}
\end{figure}

\begin{figure}
\begin{center}
\includegraphics[width=3in,height=5in,angle=90]{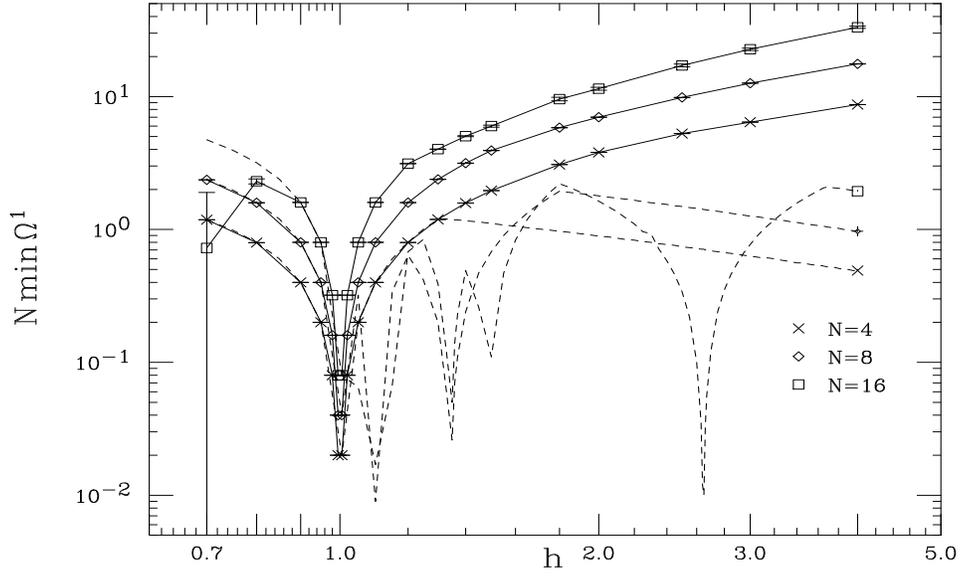}
\caption{\small The lower bound on the charged mirror inverse propagator eigenvalue,  $N {\rm min}\; \Omega^1$, for $\kappa = 0.5$. The dashed lines show the same quantities in the broken ($\kappa \rightarrow \infty$) phase; their oscillations and dips are explained by eqn.~(\ref{smalleigenvalues}). The minimum mass eigenvalues   are obtained, in units of $L_{phys.}^{-1}$, by multiplying the plotted quantity by the dimensionless (large) Yukawa coupling $Y$; see Section \ref{chiralmassless}.}
\label{fig:om1_05} 
\end{center}
\end{figure}

\begin{figure}
\begin{center}
\includegraphics[width=3in,height=5in,angle=90]{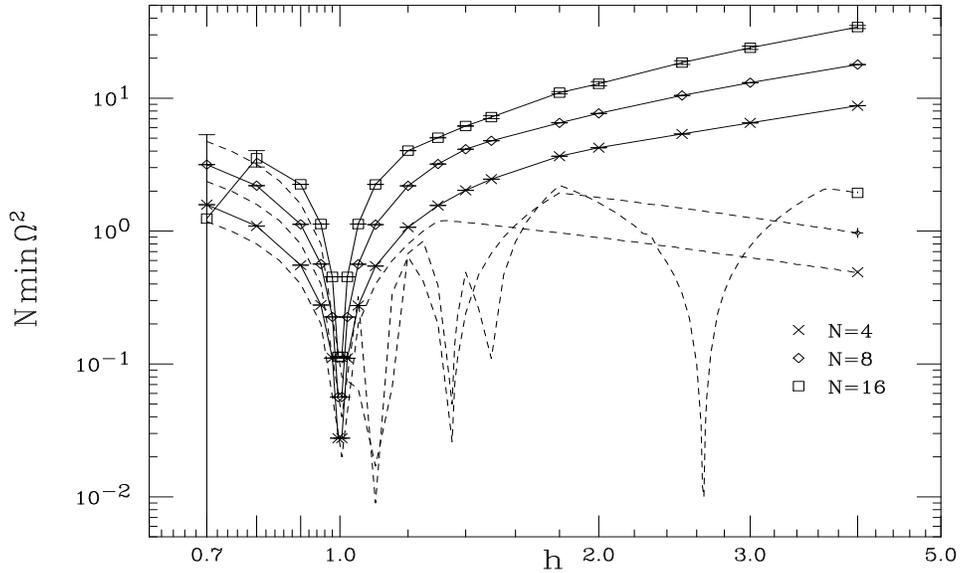}
\caption{\small The lower bound on the neutral mirror fermion eigenvalue,  $N {\rm min}\; \Omega^2$, for $\kappa = 0.5$. Note that   the charged fermions are heavier than the neutral in the symmetric phase.}
\label{fig:om2_05} 
\end{center}
\end{figure}

\begin{figure}
\begin{center}
\includegraphics[width=3in,height=5in,angle=90]{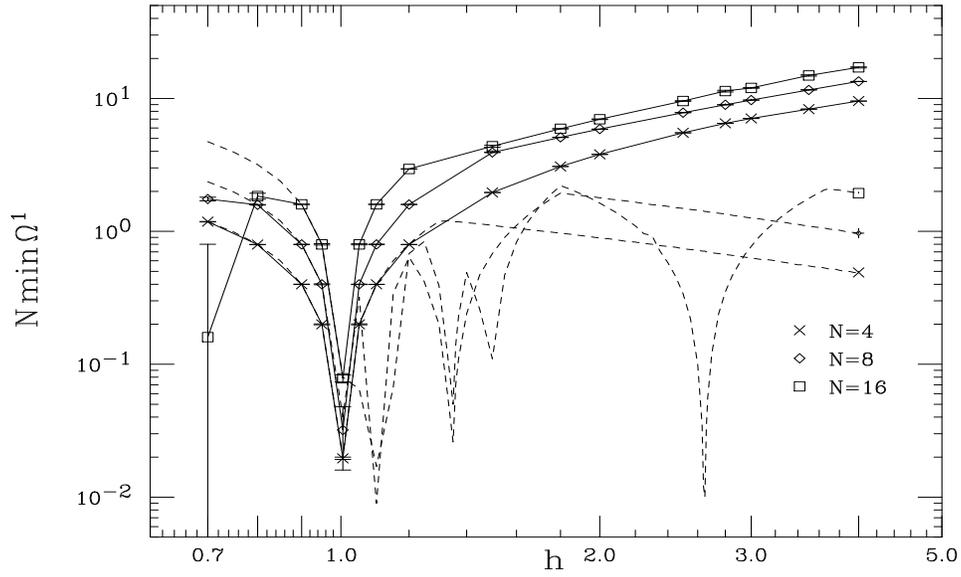}
\caption{\small The lower bound on the charged mirror fermion eigenvalue, $N {\rm min}\; \Omega^1$, for $\kappa = 0.1$. }
\label{fig:om1_01} 
\end{center}
\end{figure}

\begin{figure}
\begin{center}
\includegraphics[width=3in,height=5in,angle=90]{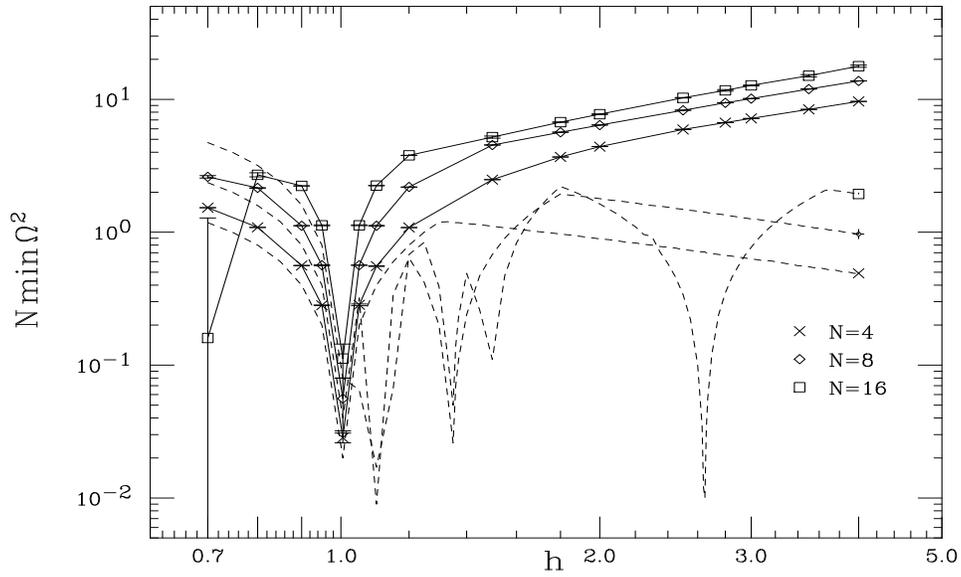}
\caption{\small The lower bound on the neutral mirror fermion eigenvalue, $N {\rm min}\; \Omega^2$ for $\kappa = 0.1$. }
\label{fig:om2_01} 
\end{center}
\end{figure}

\end{document}